\documentclass[manuscript,screen,acmtog]{acmart}

\usepackage{color}
\usepackage{rotate}
\usepackage{rotating}
\usepackage{graphicx}
\usepackage{array}
\usepackage{amsmath}

\definecolor{orange}{rgb}{1,0.5,0}

\AtBeginDocument{%
  }

\setcopyright{acmlicensed}
\copyrightyear{2025}
\acmYear{2025}
\acmDOI{XXXXXXX.XXXXXXX}
\acmISBN{978-1-4503-XXXX-X/18/06}

\acmSubmissionID{1282}

\begin{document}

\title{ChildlikeSHAPES: Semantic Hierarchical Region Parsing for Animating Figure Drawings}
\author{Astitva Srivastava}
\authornote{Equal Contribution.}
\authornote{Current affiliation: IIIT Hyderabad, Hyderabad, India }
\affiliation{%
  \institution{Meta}
  \state{CA}
  \city{Sausalito}
  \country{USA}
}
\email{astitva.srivastava@research.iiit.ac.in}

\author{Harrison Jesse Smith}
\authornotemark[1]
\authornote{Corresponding author.}
\affiliation{%
  \institution{Meta}
  \city{Sausalito}
  \state{CA}
  \country{USA}
}
\email{hjessmith@gmail.com}

\author{Thu Nguyen-Phuoc}
\affiliation{%
  \institution{Meta}
  \city{London}
  \country{UK}
}
\email{thu.h.nguyen.phuoc@gmail.com}

\author{Yuting Ye}
\affiliation{%
  \institution{Meta}
  \city{Redmond}
  \state{WA}
  \country{USA}
}
\email{yuting.ye@gmail.com}






\begin{abstract}
Childlike human figure drawings represent one of humanity's most accessible forms of character expression, yet automatically analyzing their contents remains a significant challenge.
While semantic segmentation of realistic humans has recently advanced considerably, existing models often fail when confronted with the abstract, representational nature of childlike drawings.
This semantic understanding is a crucial prerequisite for animation tools that seek to modify figures while preserving their unique style.
To help achieve this, we propose a novel hierarchical segmentation model, built upon the architecture and pre-trained SAM, to quickly and accurately obtain these semantic labels. 
Our model achieves higher accuracy than state-of-the-art segmentation models focused on realistic humans and cartoon figures, even after fine-tuning.
We demonstrate the value of our model for semantic segmentation through multiple applications: a fully automatic facial animation pipeline, a figure relighting pipeline,
improvements to an existing childlike human figure drawing animation method, and generalization to out-of-domain figures.
Finally, to support future work in this area, we introduce a dataset of 16,000 childlike drawings with pixel-level annotations across 25 semantic categories. 
Our work can enable entirely new, easily accessible tools for hand-drawn character animation, and our dataset can enable new lines of inquiry in a variety of graphics and human-centric research fields.
\end{abstract}





\begin{teaserfigure}
  \centering
  \includegraphics[width=\textwidth]{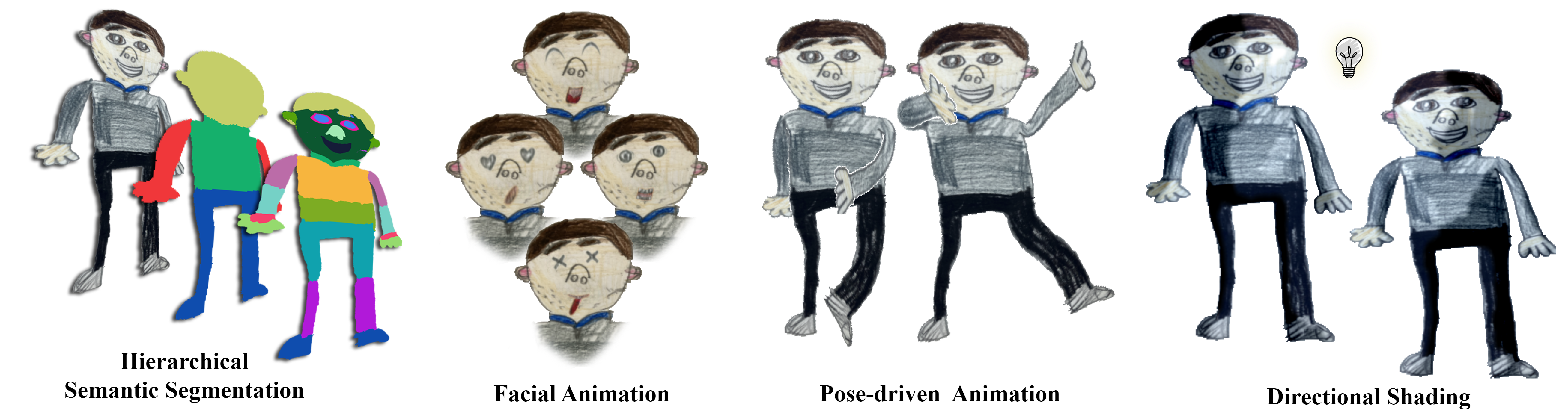}
  \caption{Our hierarchical semantic segmentation model, trained on our ChildlikeSHAPES dataset, enables various downstream applications such as semantics-guided body \& facial animation, and hand-drawn character shading.}
  \Description{SHAPES}
  \label{fig:teaser}
\end{teaserfigure}

\maketitle

\section{Introduction}

Semantic segmentation had advanced dramatically in recent years, with models now approaching human-level performance on photographs~\cite{sapien,cheng2021mask2former} . Yet a striking gap remains between human and machine perception: while we effortlessly understand abstract representations, even state-of-the-art models struggle with seemingly simple drawings. This is particularly evident when they are confronted with \textit{childlike drawings}, the abstract, highly-stylized, representational depictions that emerge early in childhood (see Sec.\ref{childlike_drawing_style}).

Within this domain, humanlike (bipedal) figures represent a particularly important and challenging case. 
Such figures exhibit extreme variability in their representation: they can have widely different proportions and asymmetries, varied connective morphologies~\cite{cox2014drawings}, and may be drawn either hollow or with interior texture. Their structure might deviate significantly from human anatomy, with unimportant details or entire body parts left absent. They may even include non-human features like wings or horns when used to represent fanciful characters, animals, or anthropomorphized objects. Though we refer to them as \textit{childlike}, these forms can be drawn by virtually anyone, regardless of age or artistic training. This makes them one of humanity's most accessible forms of character representation, and the ability to easily animate these figures would empower everyone, regardless of training or technical expertise, with a new, uniquely personal form of creative expression.

Automatically understanding the semantic structure of these figures is a crucial first step towards this vision. However, their abstract nature poses unique challenges for computer vision models, as they rely on simple regions whose meaning depends heavily on context, lacking the texture, color, and realistic proportions that modern segmentation models typically use. 

We address these challenges by developing \textit{CharSegNet}, a hierarchical semantic segmentation model that first understands broader figure structure before analyzing detailed features. Training on our novel ChildlikeSHAPES dataset of 16,075 manually annotated drawings and leveraging a fine-tuned encoder, our model learns to understand these figures through their geometric relationships rather than surface-level features. This geometric understanding enables remarkable generalization - our model can effectively process figures from cave paintings to street signs, and can be adapted to new domains like cartoon animals with minimal retraining. We demonstrate the value of semantic understanding through several animation applications: automatic facial animation and lip syncing, figure shading, and improving body animation.

Our primary contributions include:
\begin{enumerate}
    \item CharSegNet, a hierarchical semantic segmentation model designed specifically for childlike drawings.
    \item ChildlikeSHAPES dataset, a diverse collection of manually annotation figure drawings that will be publicly released.
    \item Extensive evaluations demonstrating both our model's effectiveness and the importance of key design decisions.
    \item Multiple novel applications showing how semantic understanding can enhance character animation tools.
\end{enumerate}

Our work also enables new forms of creative expression and sheds light on universal patterns to represent human figures in drawings.


\section{Background}
We provide background in three key areas to contextualize our work: the nature and characteristics of childlike drawings, recent advances in semantic segmentation, and approaches to 2D figure animation across traditional and learning-based methods.
\subsection{Childlike Drawings}
\label{childlike_drawing_style}

In the context of this work, we define \textit{childlike drawings} to be drawings exhibiting the schematic or pre-schematic representational style that typically emerges around 4-6 years of age \cite{lowenfeld1957creative}.
This style emerges following two developmental milestones: 
first, the cultivation of hand strength and coordination sufficient to control a marker; second, the realization that their own markings can resemble real-world objects \cite{luquet1913dessins}.
Although it first appears early in life, this style of drawing remains accessible well beyond childhood; adults with no artistic background, when called upon to draw, might well use this childlike style.
As such, childlike drawing represents one of the most widely accessible forms of visual communication.

Childlike drawings differ fundamentally from realistic images and lifelike drawings due to both the artists' technical limitations and their intentions.
These drawings are created without formal techniques such as perspective construction, proportional accuracy, and surface rendering, making visually accurate depiction impossible. 
However, the goal of these drawings is not visual accuracy, but recognizability; artists focus on capturing essential details that make their subjects recognizable to a viewer \cite{jolley1991children}.

To do this, childlike drawings use \textit{regions} as their fundamental building blocks \cite{willats2006making}.
These regions act as visual metaphors for 3D objects, with their shape and placement determined by the important details of the subject they represent.
Long objects, like arms, become elongated regions, while a squat object, like a head or torso, may become circular regions. 
Understanding what a region represents requires considering its context within the entire character.
A single line, for instance, can serve as a hair strand, mouth, leg, or torso.
Similarly, a dot can serve as an eye, freckle, or decorative mark depending upon the context of the character.

To faithfully animate these drawings, we must first identify what each region represents.
This semantic understanding allows us to animate the character by modifying its regions in ways that change viewer perception while maintaining the original artistic style.
Animation systems that ignore semantic region understanding risk compromising both the perceptual clarity and essential characteristics that make these drawings so unique.

\subsection{Semantic Segmentation}
Semantic segmentation is a long-standing problem in computer vision with numerous downstream applications, including animation. Recent advancements in foundation vision-language models \cite{pmlr-v139-radford21a} have enabled the development of text-based segmentation methods \cite{lueddecke22_cvpr, PNVR_2023_ICCV}. However, these approaches are not precise and struggle with the domain gap between photorealistic and non-photorealistic domains \cite{cheng2021mask2former}.
More specialized segmentation methods focus primarily on photorealistic human subjects~\cite{sapien} and non-photorealistic characters, such as anime and cartoon images~\cite{CPNet, wan2020dense, Wu_cartoonImage, 10.1145/3664647.3680879}. CPNet~\cite{CPNet} presents a method for parsing cartoon drawings. Additionally, their dataset is not public and does not contain fine-grained facial labels. ~\citet{Wu_cartoonImage} uses superpixels to perform part-level segmentation but does not include any semantic labels. Meanwhile, method by~\citet{wan2020dense} focuses solely on semantic segmentation for cartoon dog images.
Non-photorealistic images, such as cartoons, offer a wider range of appearances and shapes compared to photorealistic images. However, they still fall short in matching the vast diversity of appearances and shapes found in childlike drawings. As a result, previous methods continue to struggle with semantic segmentation in childlike drawings. \citet{yaniv_2019} tackles a very relevant task to ours: detecting facial landmarks in portrait paintings, from abstract to photorealistic. However, their work only focuses on faces, and for downstream applications such as animations, landmarks cannot be used to create animations that go beyond geometric warping.

\subsection{Animating 2D Subjects}
Converting static 2D characters into animated figures has been extensively studied in computer graphics, mainly using mesh deformation techniques.
Many approaches build upon Igarashi et al.'s as-rigid-as-possible (ARAP) shape manipulation~\cite{10.1145/1073204.1073323}. It has been extended to support depth-based scaling for 3D-to-2D motion retargeting~\cite{Hornung2007anim2Dpicmotion} and sketch stroke preservation~\cite{10.1145/3173574.3174236}. More recently, ~\citet{10.1145/3592788} applied it to automatically animate photographs of childlike hand drawings. However, these methods rely on foreground masks, instead of semantic information, to build their meshes.

An alternative approach creates 3D models from 2D inputs for animation. \textit{Photo Wake Up}~\cite{weng2019photo} and \textit{DrawingSpinUp}~\cite{10.1145/3680528.3687593} generate rigged and textured 3D meshes from single photographs, while \textit{Sketch2Mesh}~\cite{guillard2021sketch2mesh} creates meshes directly from digital sketches. Although it is straightforward to animate the resulting 3D models, this approach often sacrifices the distinctive 2D feel of hand drawings.

Recent generative image and video models have demonstrated impressive results in generating animations from a single input image~\cite{X-portrait, followEmoji} or additionally an input audio~\cite{emote, xu2024vasa, wei2024aniportrait}. However, most of these models focus on photorealistic human animation and struggle with hand-drawn images that deviate significantly from realistic representations. Some recent generative methods explore animation of more abstract subjects~\cite{Gal_2024_CVPR,wu2024aniclipart}, but their focus on vectorized content precludes easy application to hand-drawn images.

Beyond fully automatic approaches, significant work has focused on making the animation process more accessible to casual users. While tools like Adobe Character Animator~\cite{character_animator} simplify the animation process, they still are not immediate enough for casual users who want to quickly animate their paper drawings. Research projects have explored more accessible interfaces, from physical paper manipulation~\cite{barnes2008video} to digital sketch-based frameworks~\cite{Dvoroznak20-SA} and kinetic textures~\cite{kazi2014draco}. Our work's semantic understanding can augment such systems, enabling more sophisticated animation while maintaining ease of use.

\section{Dataset}
While humans easily identify the semantic parts of hand-drawn characters, models trained on photographs of realistic objects and high-quality vector art often struggle with such drawings.
To help address this limitation, we present a large-scale semantic segmentation dataset of 16,075 images of childlike drawings.
Each image is manually annotated with pixel-level labels for 25 semantic parts, providing a valuable resource for understanding how human-like characters are represented in childlike drawings.

Our dataset builds upon the Amateur Drawings Dataset~\cite{10.1145/3592788}, a collection of childlike drawing images submitted anonymously by users to an online demo. From this source, we randomly selected and cropped 20,400 images using the included character bounding box annotations. Each cropped image was manually reviewed to ensure that it contained exactly one full-bodied bipedal character; images that were too dark, had low contrast, or were too blurry for accurate segmentation were also excluded. This process yielded 16,075 images (78.8\%) suitable for inclusion in the dataset.

The annotation process was designed to capture the semantic structure of childlike drawings while addressing the substantial variability and inherent ambiguity of the domain.
We defined 25 semantic part categories, organized broadly into body and facial features (see Appendix A.1 for full category list).
These categories strike a balance between granularity and practicality; frequently occurring parts in childlike drawings~\cite{goodenough1926measurement} received dedicated labels, while less common or idiosyncratic elements were grouped under the umbrella label "accessory".

\begin{figure*}
    \centering
    \includegraphics[width=\textwidth]{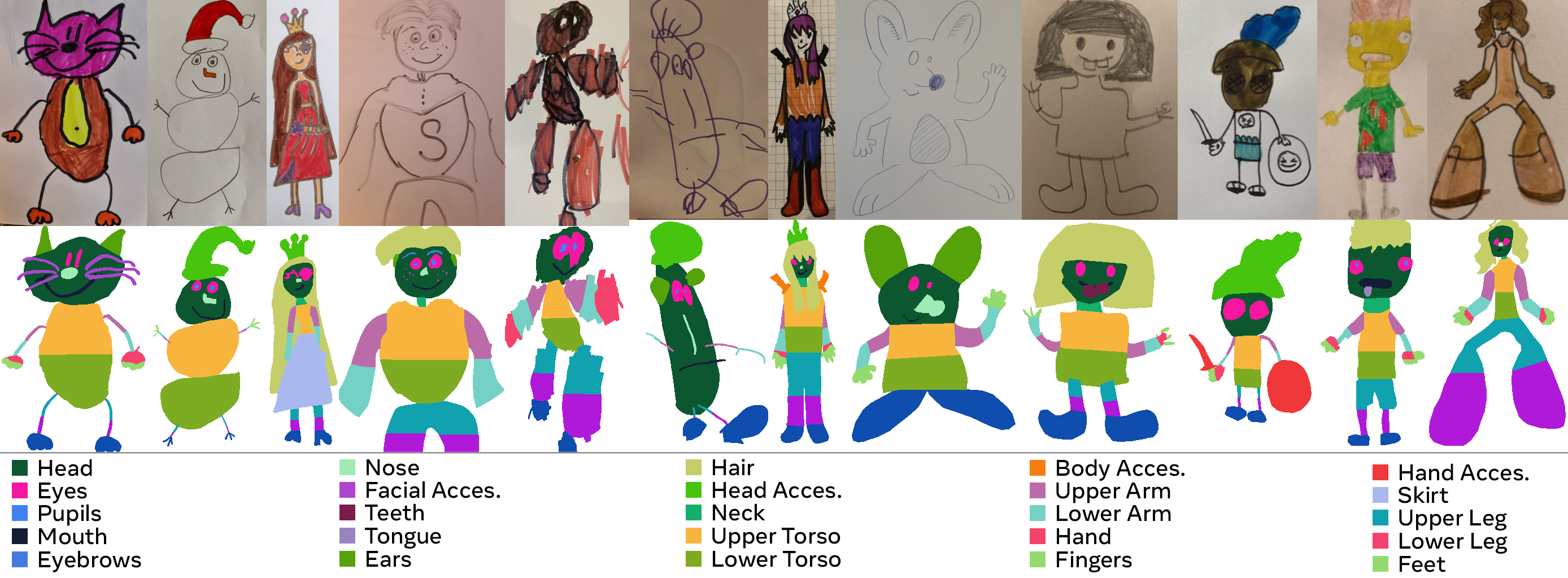}
    \caption{Representative images and manually created annotations for the 25 classes in our ChildlikeSHAPES dataset.}
    \label{fig:dataset_examples}
\end{figure*}

A team of professional annotators created the annotations using a custom browser-based tool. 
Each annotator completed initial training to ensure familiarity with both the tool and annotation guidelines (See Appendix A.1).
Quality assurance involved multiple stages: senior team members regularly reviewed annotators' work, all annotations underwent a second pass to verify guideline adherence, and the research team conducted random audits.
This rigorous process was essential given the inherent ambiguity and subjectivity in interpreting amateur drawings.
We show examples of the resulting annotations in Fig.\ref{fig:dataset_examples}.

The final dataset comprises 14,075 training images and 2,000 test images, with each image retaining its original uniquely identifying name from the Amateur Drawings Dataset. Image resolutions vary, reflecting the diverse nature of the source.  We provide a table comparing our dataset against similar ones in Appendix A.2. To provide insights into the characteristics of the dataset, we provide part-level summary statistics in Appendix A.3. We plan to release the complete dataset, including all images and corresponding semantic segmentation masks, upon acceptance of the paper.

\section{Method}

\begin{figure*}
    \centering
    \includegraphics[width=\linewidth]{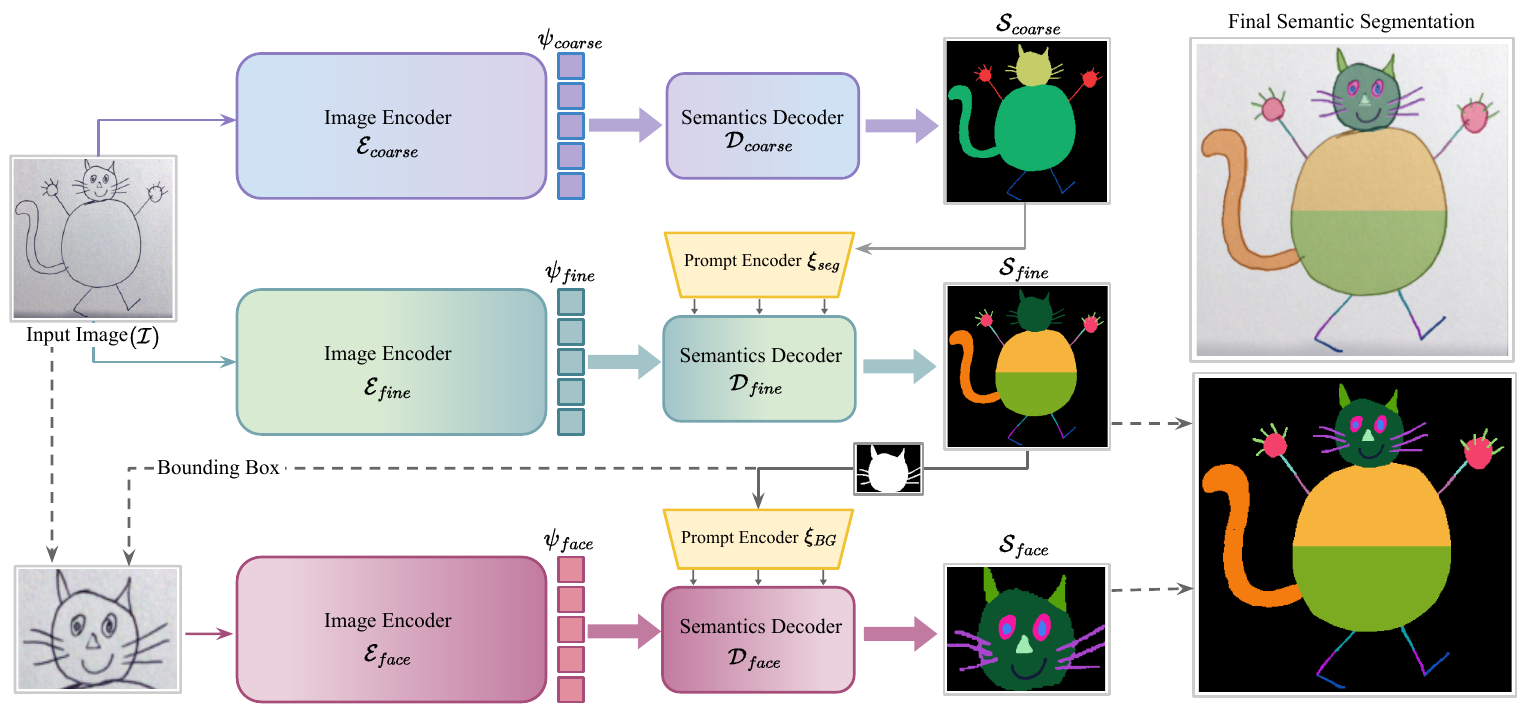}
    \caption{Proposed pipeline for Hierarchical Semantic Segmentation. The input image is passed to the coarse segmentation network, which predicts a coarse semantic mask. The coarse mask is then fed to the fine network via the prompt encoder to predict part-level semantic segmentation. Finally, the face region from the fine mask is cropped and passed to the face-segmentation network to predict detailed semantics in the face region.}
    \label{fig:architecture_c2f}
\end{figure*}

Our semantic segmentation approach leverages key insights about how childlike drawings represent figures through abstract regions.
CharSegNet employs a hierarchical, coarse-to-fine prediction strategy, as illustrated in Fig.~\ref{fig:architecture_c2f}, rather than identifying all semantic regions simultaneously.
By first understanding the broader structure before analyzing details, CharSegNet resolves ambiguous regions whose semantic meaning depends on their context with the figure.


\subsubsection{Architecture}
Our approach adapts the Segment Anything Model (SAM)~\cite{kirillov2023segany}, a task-agnostic segmentation model, to accommodate our hierarchical prediction design.
SAM's architecture offers several key advantages: its ViT-based encoder provides multi-scale region understanding crucial for processing diverse drawing styles, while its prompt encoder enables flexible incorporation of additional information during prediction. This flexibility is particularly valuable for our coarse-to-fine approach, enabling the use of coarse predictions as input for finer detail prediction.

We modify SAM's architecture for our specific needs. We initialize the image encoder and mask decoder with SAM's pre-trained weights, while randomly re-initializing the prompt encoder weights to adapt them to our hierarchical prediction task.
We repurpose the prompt encoder, originally designed to handle various types of user input, to process semantic mask information between prediction stages, leveraging SAM's general segmentation capabilities while specializing for our specific task.

\subsubsection{Hierarchical Semantic Segmentation} 
Our hierarchical approach processes the input image $\mathcal{I}$ in three stages (Fig.~\ref{fig:architecture_c2f}).
First, a coarse prediction stage (top) uses encoder $\mathcal{E}_{coarse}$ to obtain image embeddings $\mathcal{\psi}_{coarse}$ and decoder $\mathcal{D}_{coarse}$ to identify four main regions within the character: head, body, arms, and legs ($\mathcal{S}_{coarse}$). 

The resulting coarse mask $\mathcal{S}_{coarse}$ then guides a second stage of fine-grained prediction (middle). Here, encoder $\mathcal{E}_{fine}$ processes  image $\mathcal{I}$ to create embedding $\mathcal{\psi}_{fine}$.
Fine semantic decoder $\mathcal{D}_{fine}$ decodes the embedding, guided by coarse segmentation predictions $\mathcal{S}_{coarse}$ as encoded by prompt encoder $\mathcal{\xi}_{seg}$. The decoder predicts $\mathcal{S}_{fine}$, a semantic mask identifying 14 different classes, corresponding of all of the final labels except those representing facial features.

Facial features require special handling due to their frequently small size in the full image. In the facial feature prediction stage (lower), we extract the facial region from $\mathcal{S}_{fine}$ and use it to crop image $\mathcal{I}$. We then feed this cropped image into face encoder $\mathcal{E}_{face}$, obtaining embedding $\mathcal{\psi}_{face}$.  Face semantic decoder $\mathcal{D}_{face}$ then decodes this embedding, guided by a face binary mask as encoded by prompt encoder $\mathcal{\xi}_{BG}$, and predicts $\mathcal{S}_{face}$, a mask identifying 11 semantic classes specific to head and face features. 

The final segmentation is obtained by compositing $\mathcal{S}_{fine}$ and $\mathcal{S}_{face}$, resulting in a 25-class segmentation (Fig.~\ref{fig:dataset_examples}).
See Appendix B. Inference for additional training details.

\section{Applications}
\label{sec:applications}
In this section, we propose several novel applications enabled by the semantic segmentation of hand-drawn characters and the proposed dataset, focusing on automating several tasks related to a conventional animation pipeline. We mainly focus on automating the tedious task of facial animation to a given audio (Sec. \ref{sec:facial_animation_lip_sync}), animating hand-drawn characters to a given human-motion sequence (Sec. \ref{sec:body_animation_improvements}), and the novel problem of non-photorealistic shading of character drawings (Sec. \ref{sec:relighting}).
\subsection{Facial Animation}
\label{sec:facial_animation_lip_sync}
\subsubsection{Facial Expression Generation}
Creating new facial expressions for hand-drawn figures traditionally requires manually drawing multiple expressions, which can be time-consuming and tedious.
Alternatively, the original figure's expression could be changed by warping the image using facial landmarks. However, automatic facial landmark identification on childlike drawings is an open and potentially ill-posed research problem.
We sidestep these issues by leveraging CharSegNet's semantic understanding to enable style-preserving expression generation through \textit{CharFaceGAN}, a deep generative model that converts semantic masks into drawings while maintaining the unique artistic style of an input reference drawing.

CharFaceGAN adopts Semantic Region-Adaptive Normalization (SEAN)~\cite{zhu2019sean}, which enables style-preserving image synthesis guided by semantic masks.
We train CharFaceGAN on cropped faces from the ChildlikeSHAPES dataset, enabling it to learn style codes that capture region-specific artistic styles in childlike drawings.
Once trained, CharFaceGAN can reconstruct a face from the StyleGAN latent space using both the semantic mask and the original face image through inversion. 
By modifying specific regions in the semantic mask, we can generate new facial features while preserving the original face's unique artistic style.

This approach enables the creation of new facial assets by modifying semantic labels rather than drawing new expressions from scratch. Generated assets maintain consistent style with the original face while allowing for diverse expressions. For example, modifying the mouth's semantic mask can create a different mouth shape. 
This process is remarkably fast, taking under $2.5$ seconds per facial set and allowing for parallel generation of multiple assets.
See the supplemental video for an example of this interactive process.
Additional implementation details can be found in Appendix B.

\subsubsection{Preset-Based Asset Generation}
While CharFaceGAN enables expression generation through semantic mask manipulation, a fully automated pipeline requires automatic mask manipulation.
Therefore, we develop a library of preset template shapes representing standard facial expressions and visemes (mouth shapes corresponding to phonemes). These presets, defined by semantic labels, can replace a figure's original expression (Fig.~\ref{fig:face_exp_gen_results}, top row).

\begin{figure*}
    \centering
    \includegraphics[width=\linewidth]{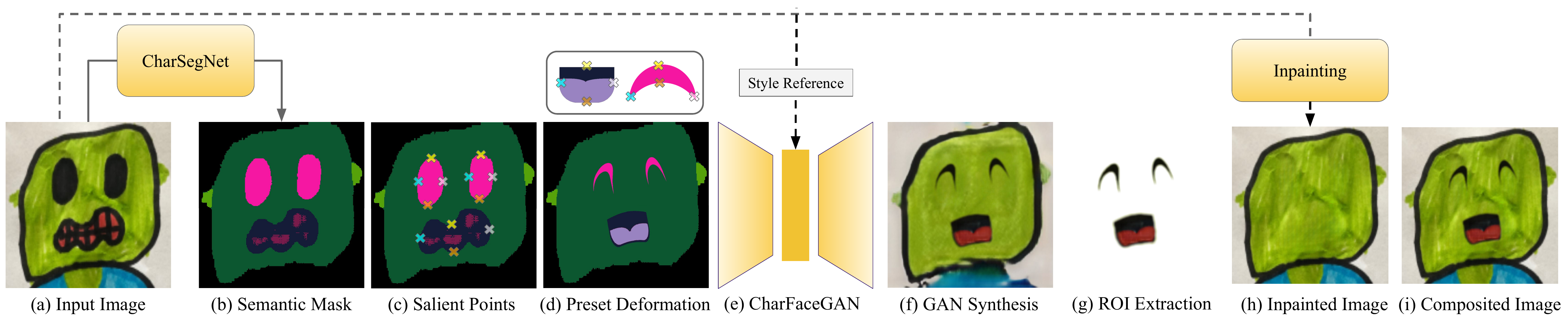}
    \caption{\textbf{Generating Novel Facial Expressions:} Proposed pipeline for preset guided novel eye \& mouth shape generation, which involves deforming the predefined semantically labeled presets via salient points, followed by a semantics-guided image synthesis to produce new details in the region of deformed presets. The newly generated details are then masked out and composited over an inpainted image (with original eyes and mouth removed) to produce new facial expressions.}
    \label{fig:preset_gen}
\end{figure*}

Our automatic expression generation pipeline is illustrated in Fig~\ref{fig:preset_gen}. Given a figure's face image and its semantic masks from CharSegNet, we first identify salient points along the contours of the mouth and eye regions. These points, marking top, bottom, left, and right boundaries, establish correspondence with predefined points on our preset shapes. These correspondences enable Thin-Plate Spline deformation~\cite{24792} to transform preset shapes from their canonical form to match the facial feature's position and proportions. The deformed preset labels then replace the original eye and mouth regions in the semantic mask, which guides CharFaceGAN in generating new facial features that maintain the character's style.

While quite close, the features generated by CharFaceGAN may not perfectly align with the semantic boundaries. To address this, we refine the semantic masks to better adhere to the generated face using the approach proposed in \citet{DBLP:journals/corr/abs-2005-02551}.
We inpaint the eye and mouth regions using LAMA~\cite{suvorov2021resolution}, which avoids unwanted feature hallucination while being fast.
The refined generated features are composited onto the inpainted image to produce the final result. As shown in Fig.~\ref{fig:face_exp_gen_results}, this pipeline can automatically generate various facial expressions by combining different eye and mouth presets while preserving the original artistic style.

\subsubsection{Audio-Driven Animation}
Building on our preset-based asset generation, we enable automatic facial animation synchronized to audio input. For lip synchronization, we first convert the audio to text and use the Montreal Forced Aligner~\cite{MFA} to obtain time-aligned phonemes. These phonemes map directly to our viseme presets, allowing us to select and composite the appropriate mouth shapes in sync with the speech. Eye animations can be controlled through various methods to specify when different expressions should appear, whether for basic actions like blinking or for conveying emotion. The result is a complete facial animation pipeline that brings hand-drawn figures to life while maintaining their original artistic style.

\subsection{Body Animation Improvements}
\label{sec:body_animation_improvements}
The semantic labels can be used to improve algorithms focused on retargeting body motion onto characters.  To demonstrate this, we identify four limitations of Smith et al.’s method \cite{10.1145/3592788}, which relies solely on character foreground masks, and demonstrate how semantic information can address these limitations (see Fig.\ref{fig:body_animation_results} and supplementary video).

\subsubsection{Foreground Segmentation} Smith et al. rely on a floodfill-based image processing technique for segmentation. They reported that it works less than 50\% of the time, with light pencil marks, unconnected lines, and lined paper, amongst the factors that commonly result in failure. We address these limitations by constructing a foreground mask from our semantic predictions. Specifically, we consider a pixel to be foreground if it belongs to any of our model's predicted semantic regions. This approach can successfully handle cases where the flood fill method fails.

\subsubsection{Part-Aware Deformation} Smith et al. deform the character mesh using as-rigid-as-possible (ARAP) mesh deformation \cite{10.1145/1073204.1073323}. However, without semantic part labeling, all edges within the mesh are given the same weight, which can lead to undesirable deformations. For example, rigid parts such as the head may warp when adjacent parts such as the arms move. We address this by incorporating semantic information into the ARAP optimization's edge weights. Let $w_{ij}$ be the weight of edge between vertices $v_i$ and $v_j$:

\begin{equation}
    w_{ij} = \alpha(L_{v_i}) + \alpha(L_{v_j})
\end{equation}

 where $L_i$ is the semantic label of the pixel closest to $v_i$ and $\alpha$ is a scaling factor based on the labels. We set $\alpha = 1$ if the label is a deformable region (arms, legs, lower torso) and $\alpha = 100$ elsewhere, resulting in a more natural and appealing deformation.

\subsubsection{Overlapping Body Parts}
Smith et al.'s method relies solely on character foreground segmentation masks, which can lead to failures when disconnected body parts overlap in the drawing. We address this limitation by leveraging semantic part labels and anatomical connectivity constraints.
Let $C(L)$ denote the set of semantic labels that can be connected to label $L$. For example,  
$C(L_{finger}) = \{L_{hand}, L_{lower\:arm}\}$.
We can effectively separate body parts that should not be connected in the following manner:
\begin{enumerate}
    \item Dilate each semantic region by one pixel to create expanded region $E_L$
    \item For each label $L$ and pixel $p \in E_L$, check if $p$ overlaps with any region $L'$ where $L' \notin C(L)$
    \item Reclassify all such overlapping pixels as background
\end{enumerate}

\subsubsection{Foot Orientation}
Feet are commonly drawn extending towards either the right or left.
Smith et al.'s method does not modify the connectivity of the character mesh, which can result in noticeable artifacts, such as the character's foot facing backwards relative to the knee's bending direction.
We address this issue by utilizing semantic labels to detect and correct the foot orientation. For each foot region, we:
\begin{enumerate}
    \item Compute its center of mass as the average position of all pixels in the region
    \item Define a reference vector $\vec{r}$ from the attached leg's knee joint to its ankle joint
    \item Classify the foot as right-facing if its center of mass lies to the right of $\vec{r}$ and left-facing otherwise
\end{enumerate}
During animation, we determine the leg's bend direction by checking whether the knee joint lies to the left or right of the vector connecting the hip and ankle joints. If the foot orientation does not match the leg's bend direction, we cut the character mesh midway between the knee and ankle, mirror the lower portion about the knee-to-ankle vector, and reconnect the meshes. 

\subsection{Non-Photoreal Shading of Hand-drawn Characters}
\label{sec:relighting}
Adding shading to hand-drawn figures traditionally requires either manual artistry or the use of 3D models with non-photorealistic rendering. While 3D approaches can provide physically accurate lighting, creating 3D models for stylized drawings is often impractical or impossible due to their abstract nature. We propose a novel approach for realistic shading without 3D information by treating shading as a conditional semantic segmentation problem. 
Our method categorizes characters into three shading levels - shadow, unlit, and highlights - based on a light source position, enabling real-time manipulation of lighting effects on hand-drawn figures.


\noindent
\textbf{\underline{SynthShade Dataset:}} To create accurate ground truth shading maps, we curate a synthetic dataset using the 3DBiCar dataset \cite{Luo_2023_CVPR} of textured 3D cartoon characters. We implement a Blender-based NPR shader to produce three types of lighting effects as discussed above. We also implement several custom stylized shaders to render flat drawing-like images of 3D cartoon meshes. We normalize all the meshes to [0,1] view-port range and randomly sample 3D positions for point light inside a $1\times1\times1$ grid around the character (excluding light positions behind the 3D character). With this approach, we create the \textbf{SynthShade} dataset of $\sim100k$ synthetic drawing-like images of cartoon-characters with accurate shading maps in $2048\times2048$ resolution. Additional implementation details can be found in Appendix B.

\noindent
\textbf{\underline{Shading as Conditional Semantic Segmentation:}} 
We repurpose CharSegNet into \textbf{CharShadeNet}, leveraging the pre-trained image encoder $\mathcal{E}_{fine}$ and introducing a new shading decoder $\mathcal{D}_{shade }$ to predict Shading Maps $\mathcal{M}_i$, conditioned on 3D point light positions $(x_i,y_i,z_i)$ encoded by a prompt encoder $\mathcal{\xi}_{light}$.
We fine-tune the entire encoder-decoder architecture using the following objective function:
\begin{equation}
    \mathcal{L}_{shade} = \lambda_{1}\mathcal{L}_{dice} + \lambda_{2}\mathcal{L}_{focal} + \lambda_{3}\mathcal{L}_{TV}
\end{equation}
where $\mathcal{L}_{TV}$ is the Total Variation loss~\cite{TVloss} as a regularizer to ensure smooth boundaries across different shading regions. During inference, CharShadeNet can take continuous lighting positions and predict shading maps with directional consistency. The predicted shading maps $\mathcal{M}_i$ are blended with the Input Image $\mathcal{I}$ to produce shaded Images $\mathcal{I}_{shade_i}$.
This demonstrates how the representation learned by SAM encoder finetuned on drawings is useful for more than just semantic segmentation tasks.
\begin{figure}
    \centering
    \includegraphics[width=\linewidth]{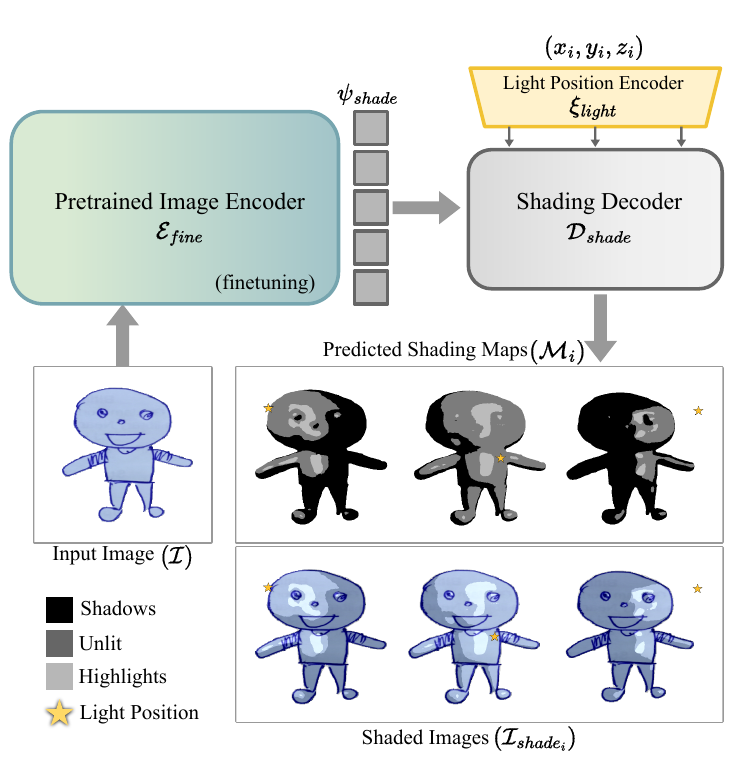}

    \caption{Proposed CharShadeNet architecture, which finetunes pretrained CharSegNet's image encoder to predict shading maps, conditioned on a 3D light position. The predicted shading are then blended with input image to enable shading. }
    \label{fig:architecture_shading}
\end{figure}

\section{Evaluation}

We evaluate CharSegNet using three standard segmentation metrics: pixel accuracy (percentage of correctly classified pixels), mean accuracy (average accuracy across all classes), and mean intersection-over-union (IoU, average overlap between predicted and ground truth regions). Our evaluation includes comparisons against state-of-the-art baselines and ablation studies of key design choices. To demonstrate CharSegNet's broader applicability, we show how it can be adapted to segment quadruped characters, achieving state-of-the-art results on cartoon dog segmentation. Finally, we analyze how CharSegNet's accuracy improves with training dataset size, providing an indication of how much data will be necessary to adapt the model to new artistic domains.
\subsubsection{Baseline Comparisons} We compare CharSegNet against three diverse baselines (Table~\ref{tab:comparisons_ablations}, top): Sapiens~\cite{sapien}, a state-of-the-art model for realistic human segmentation; DFPnet~\cite{wan2020dense}, specialized for cartoon figure segmentation; and Mask2Former~\cite{cheng2021mask2former}, a general-purpose transformer-based segmentation model. After fine-tuning all models on the ChildlikeSHAPES dataset until convergence, CharSegNet achieves superior performance across all three metrics, with improvements of $2.85\%$, $8.52\%$, and $8.12\%$ in pixel accuracy, mean accuracy, and mean IoU, respectively.

\subsubsection{Ablation Studies} Our ablation studies (Table~\ref{tab:comparisons_ablations}, bottom) examine two key components of CharSegNet. First, we demonstrate the value of hierarchical prediction by comparing against a single-stage model that predicts all semantic labels simultaneously. The hierarchical approach significantly improves accuracy across all metrics, confirming the importance of considering broader character context before predicting fine-grained features. Additionally, we evaluate the impact of fine-tuning SAM's encoder. While the pre-trained encoder provides a strong foundation for general segmentation, its features are optimized for photographs. Fine-tuning the encoder for childlike drawings yields substantial improvements, supporting our end-to-end training strategy.

\subsubsection{Cross-Domain Adaptation}
\begin{figure}
    \centering
    \includegraphics[width=\linewidth]{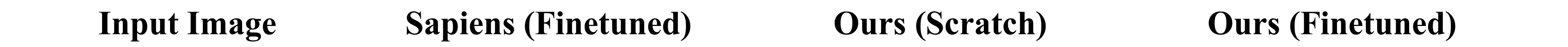}
    \includegraphics[width=\linewidth]{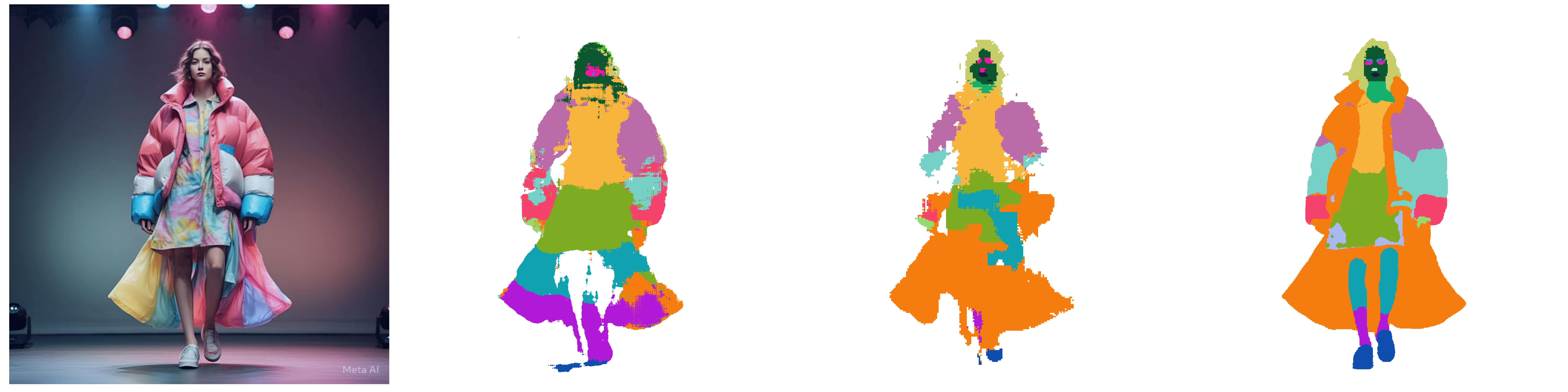}
    \includegraphics[width=\linewidth]{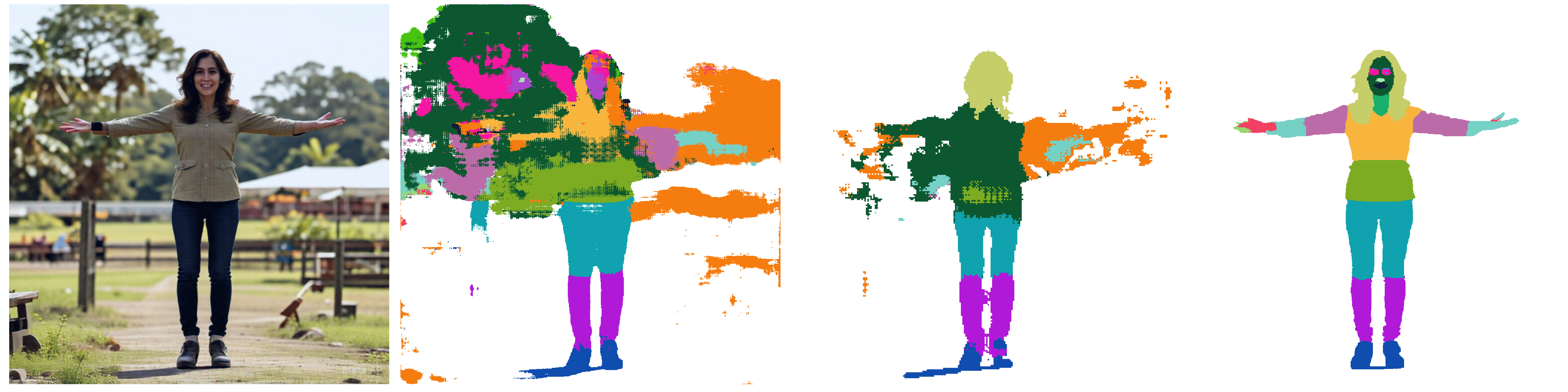}
    \includegraphics[width=\linewidth]{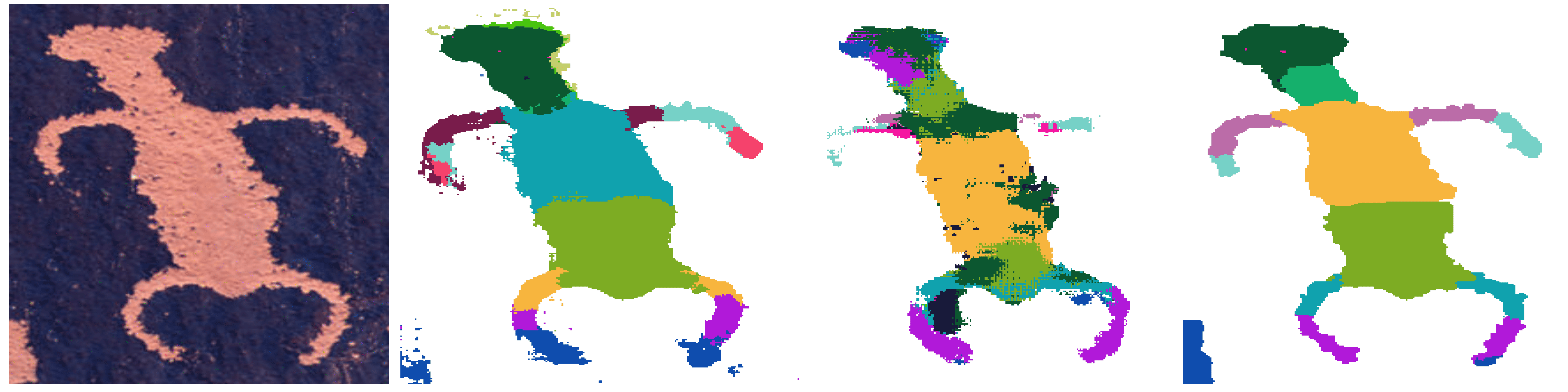}
    \includegraphics[width=\linewidth]{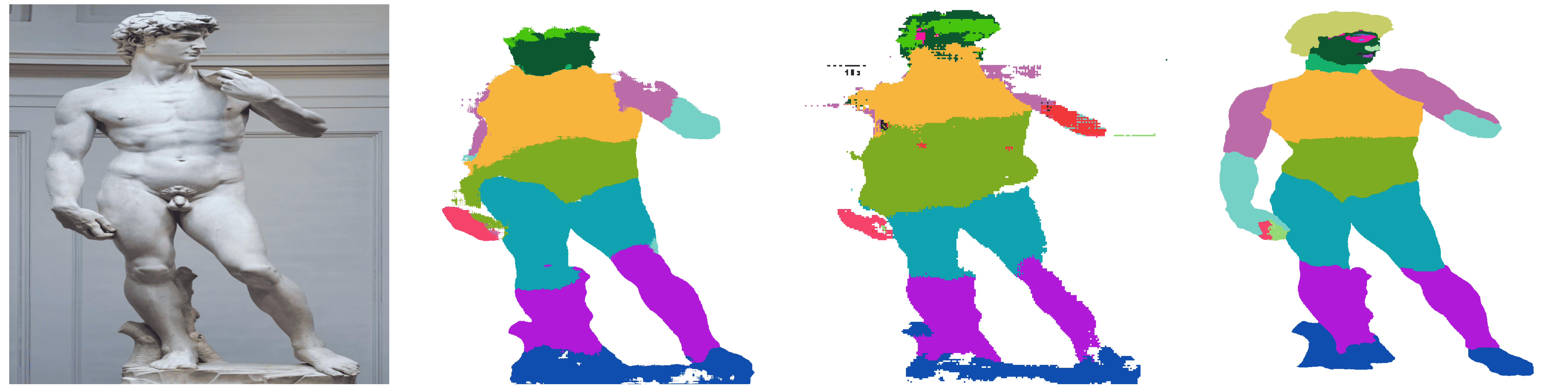}
    \caption{Comparing segmentation performance on out-of-domain samples between Sapiens~\cite{sapien} (fine-tuned), our model without SAM's pre-trained weights, and our full model.}
    \label{fig:comparison_ood}
\end{figure}

We further demonstrate CharSegNet's flexibility by adapting it to segment figures from the Cartoon Dog dataset~\cite{wan2020dense}. Because this dataset contains only seven semantic labels and no facial features, we only use CharSegNet's coarse-level segmentation module. We compare our approach against DFPnet~\cite{wan2020dense} and CartoonNet~\cite{10.1145/3664647.3680879}, evaluating two variants of CharSegNet: one trained starting from SAM's pre-trained weights, and another initialized with encoder weights that were first fine-tuned on ChildlikeSHAPES (see Table~\ref{tab:cartoon_dog}). While direct training produces reasonable results, starting from our childlike drawing-trained encoder yields
significantly better performance, achieving state-of-the-art results on cartoon dog segmentation. This demonstrates that the features learned from childlike drawings transfer effectively to other domains, a finding further supported by convincing predictions on a menagerie of out-of-domain (OOD) images (Figs.~\ref{fig:comparison_ood} and ~\ref{fig:out_of_domain}).

\subsubsection{Dataset Size Analysis}

\begin{figure}[h]
    \centering
    \includegraphics[width=\linewidth]{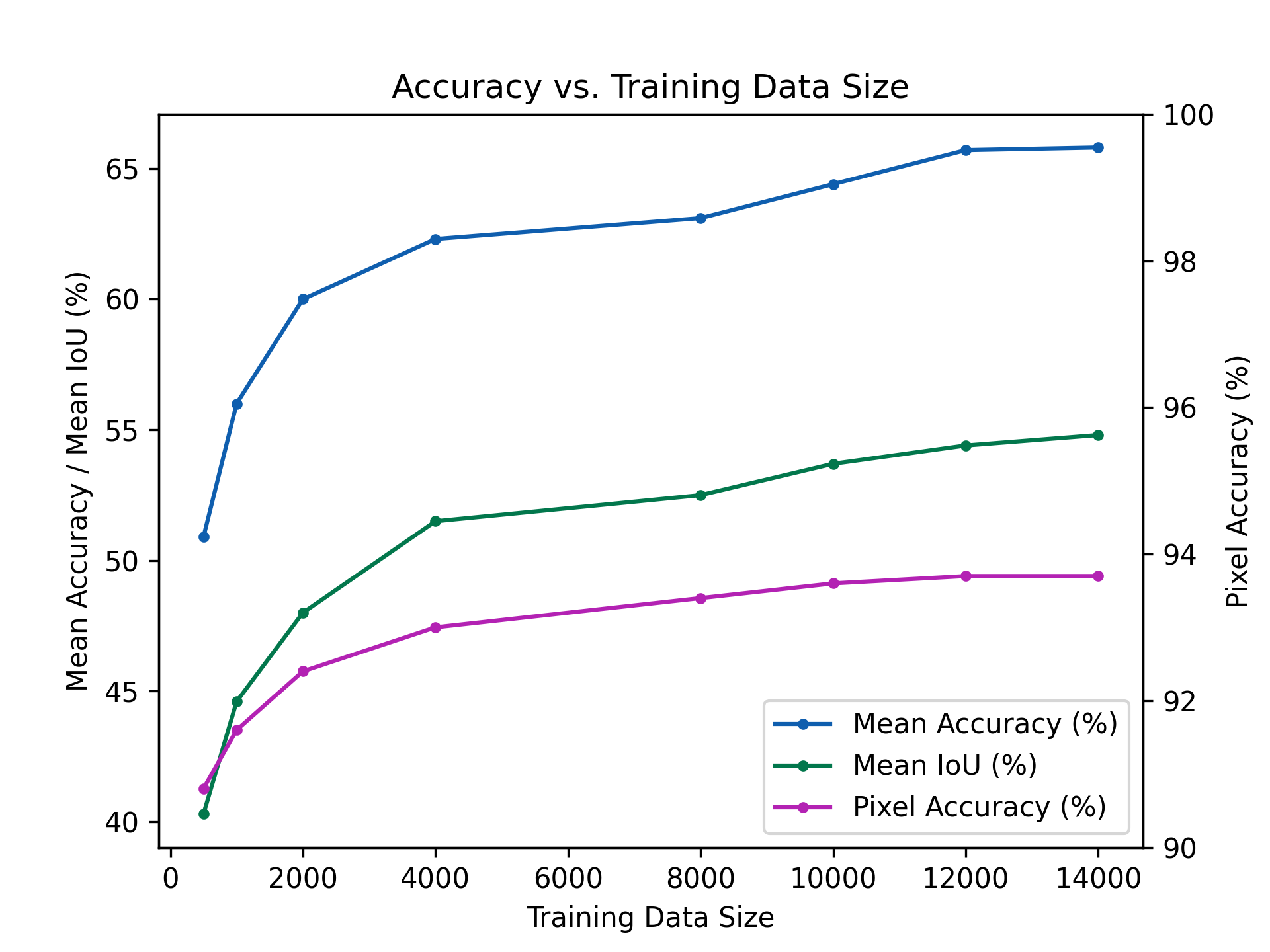}
    \caption{Relationship between training dataset size and performance metrics: pixel accuracy, mean accuracy, and mean intersection-over-union (IoU).}
    \label{fig:size_vs._accuracy}
\end{figure}

\begin{figure*}
    \centering
    \includegraphics[width=\linewidth]{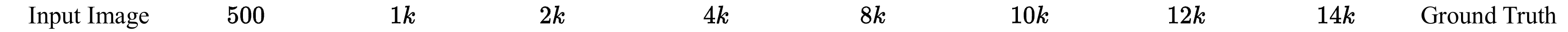}
    \includegraphics[width=\linewidth]{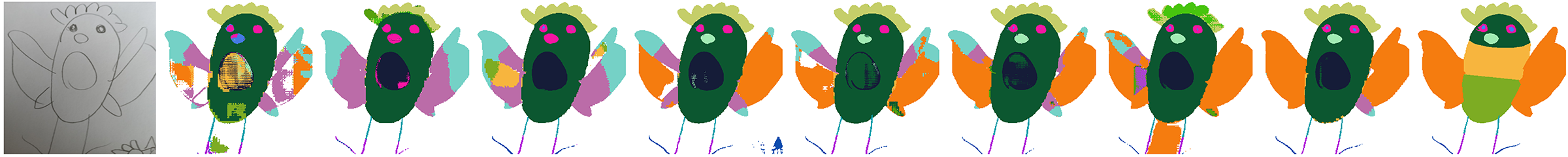}
    \includegraphics[width=\linewidth]{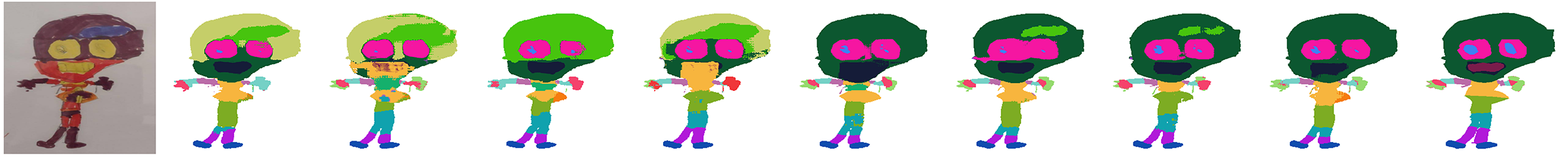}
    \includegraphics[width=\linewidth]{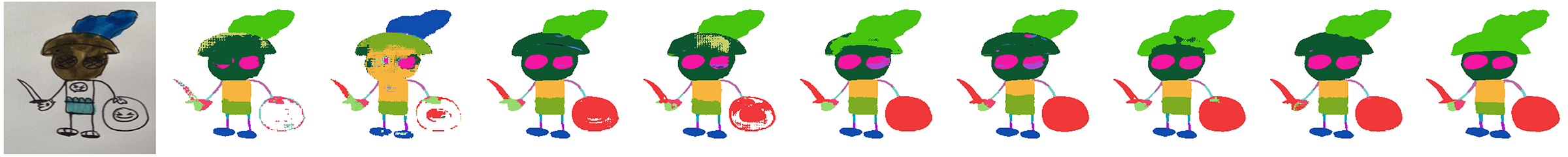}
    \includegraphics[width=\linewidth]{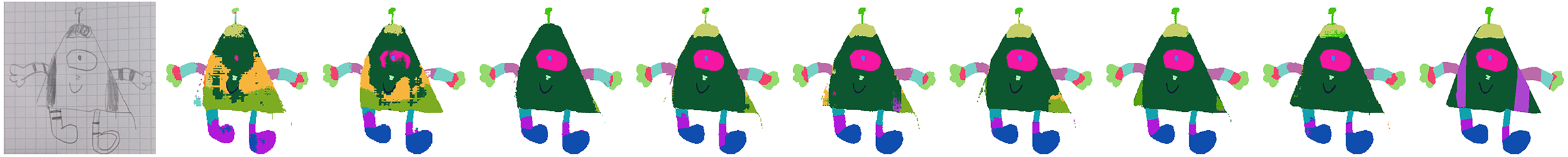}
    \caption{Segmentation results by CharSegNet trained on different training dataset sizes.}
    \label{fig:data_split_ablate}
\end{figure*}

\begin{figure*}
    \centering
    \includegraphics[width=\linewidth]{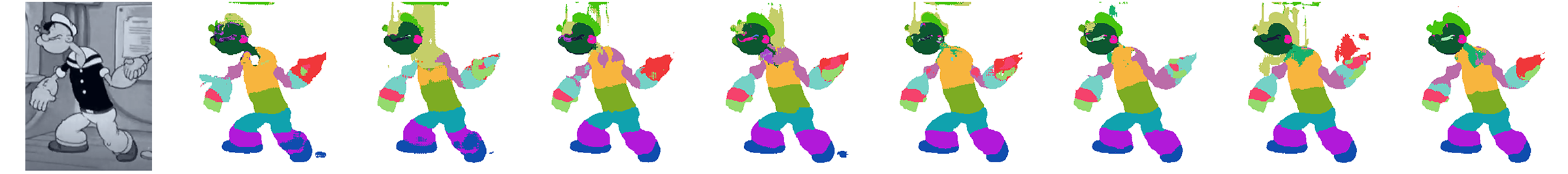}
    \includegraphics[width=\linewidth]{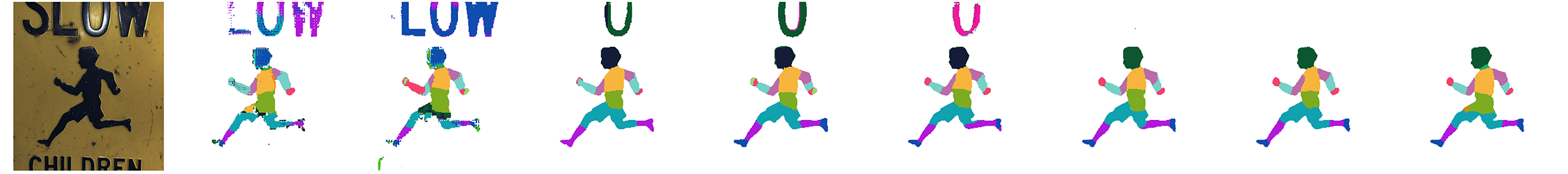}
    \includegraphics[width=\linewidth]{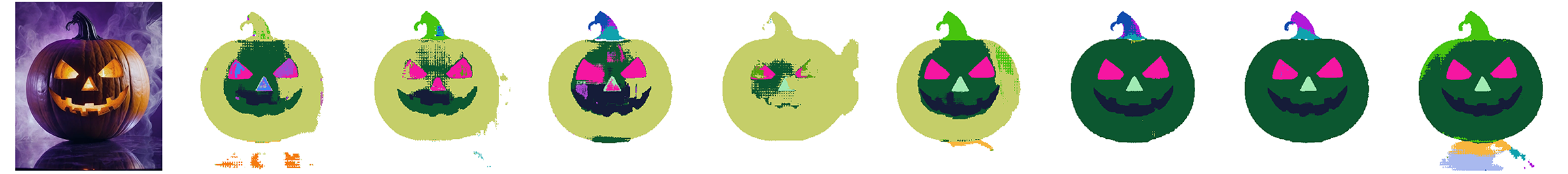}
    caption{Segmentation results for Out-of-Distribution (OOD) samples using CharSegNet trained on different training dataset sizes.}
    \label{fig:data_split_ood}
\end{figure*}

We present results on the trade-off between training sample size and accuracy (see Fig.~\ref{fig:size_vs._accuracy} and Table~\ref{tab:size_vs._accuracy}), facilitating the application of CharSegNet to other abstract domains or representational forms.
We show in-domain and OOD examples in Figs.~\ref{fig:data_split_ablate} and \ref{fig:data_split_ood}. We believe these findings will assist others in optimizing their data collection strategies.

Significant accuracy improvements are achieved with as few as 500 samples, and continue markedly until around 4,000 samples. Beyond this point, accuracy improves more gradually with additional training samples.

\begin{table}[h]
    \centering
    \caption{Accuracy Metrics by Training Data Size}
    \scalebox{0.92}{
    \begin{tabular}{l c c c c c c c c}
        \toprule
        Metric & 0.5k & 1k & 2k & 4k & 8k & 10k & 12k & 14k \\ 
        \midrule
        Pixel Acc. (\%) & 90.8 & 91.6 & 92.4 & 93.0 & 93.4 & 93.6 & 93.7 & 93.7 \\ 
        Mean Acc. (\%) & 50.9 & 56.0 & 60.0 & 62.3 & 63.1 & 64.4 & 65.7 & 65.8 \\ 
        Mean IoU (\%) & 40.3 & 44.6 & 48.0 & 51.5 & 52.5 & 53.7 & 54.4 & 54.8 \\ 
        \bottomrule
    \end{tabular}
    \label{tab:size_vs._accuracy}
    }
\end{table}

\begin{figure*}
    \centering
    \includegraphics[width=\linewidth]{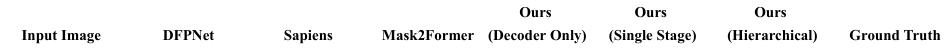}
    \includegraphics[width=\linewidth]{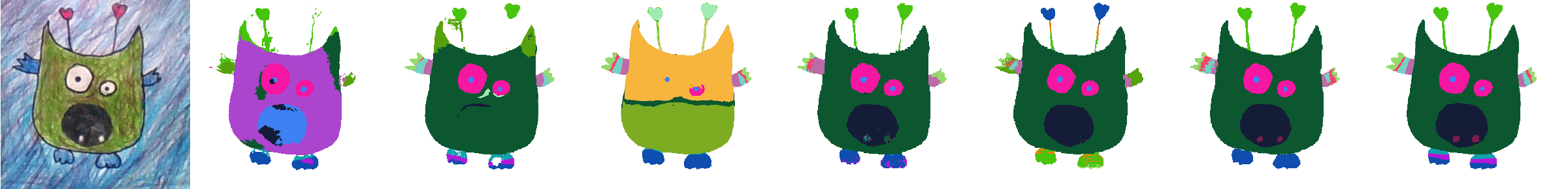}
    \includegraphics[width=\linewidth]{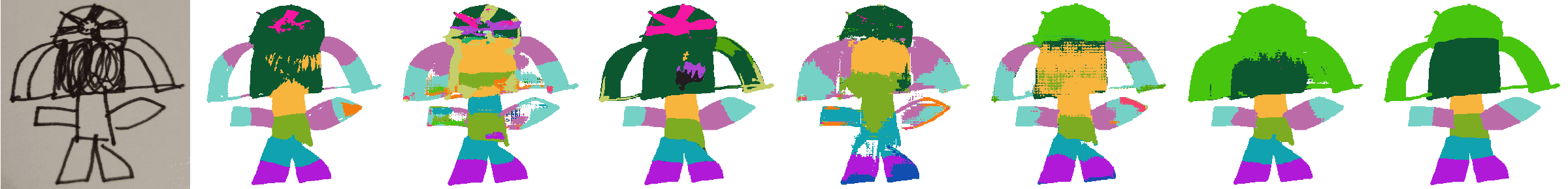}
    \includegraphics[width=\linewidth]{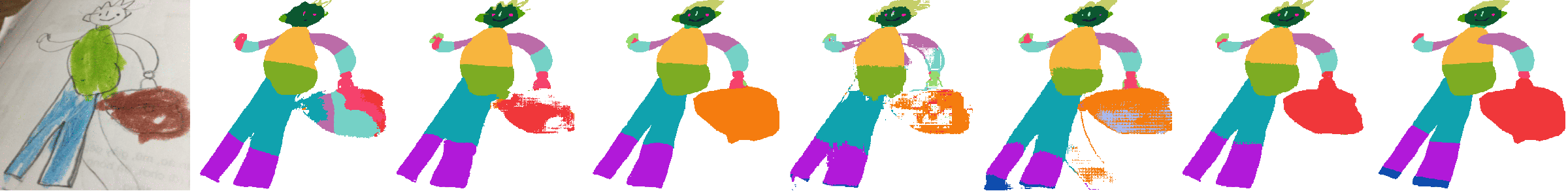}
    \includegraphics[width=\linewidth]{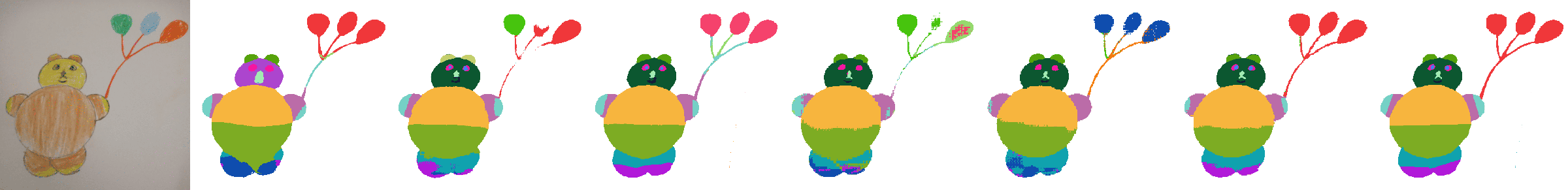}
    \caption{Qualitative comparison of segmentation results.}
    \label{fig:comparison_seg_all}
\end{figure*}

\begin{figure}
    \centering
    \includegraphics[width=\linewidth, trim={16cm 0 0 0},  clip]{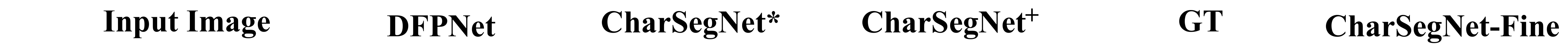}
    \includegraphics[width=\linewidth, trim={12.5cm 0 0 0},  clip]{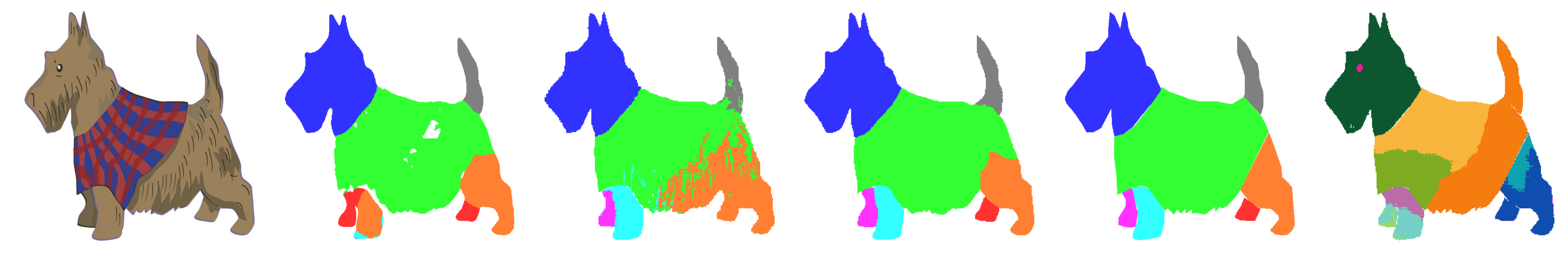}
    \includegraphics[width=\linewidth, trim={12.5cm 0 0 0},  clip]{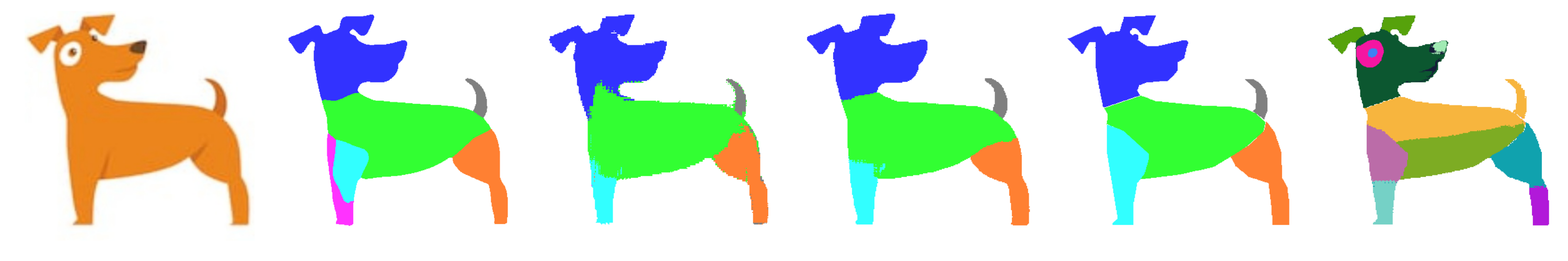}
    \includegraphics[width=\linewidth, trim={12.5cm 0 0 0},  clip]{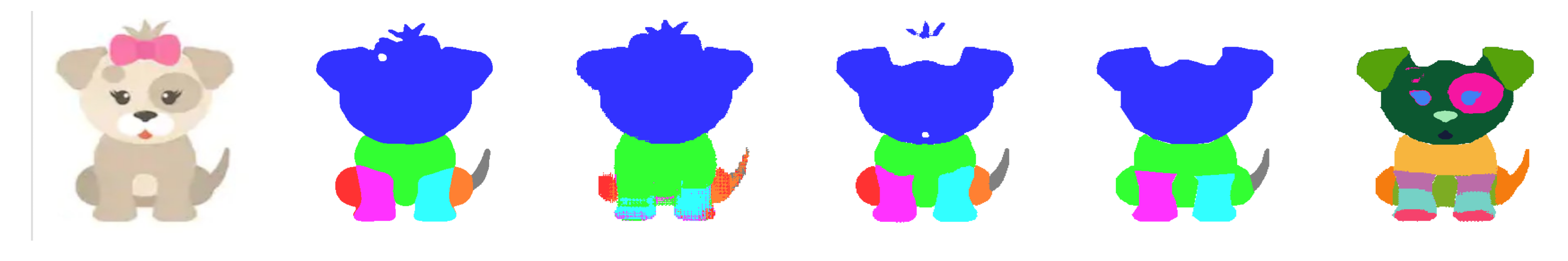}
    \includegraphics[width=\linewidth, trim={12.5cm 0 0 0},  clip]{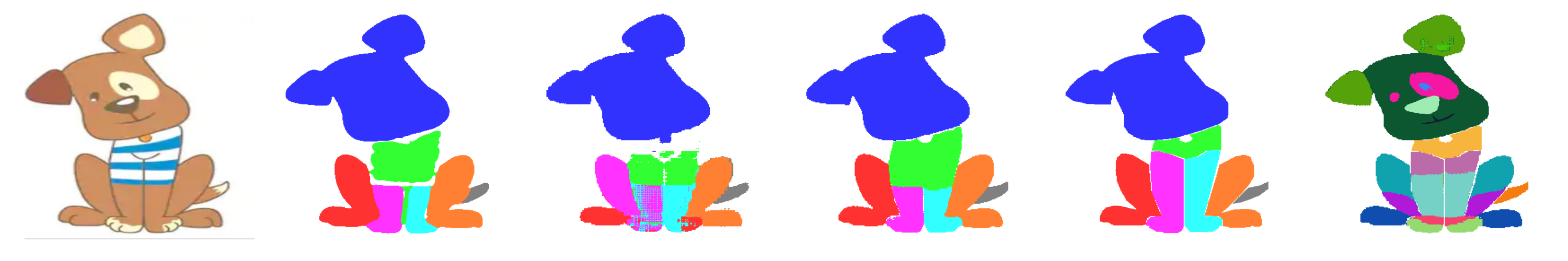}
    \includegraphics[width=\linewidth, trim={12.5cm 0 0 0},  clip]{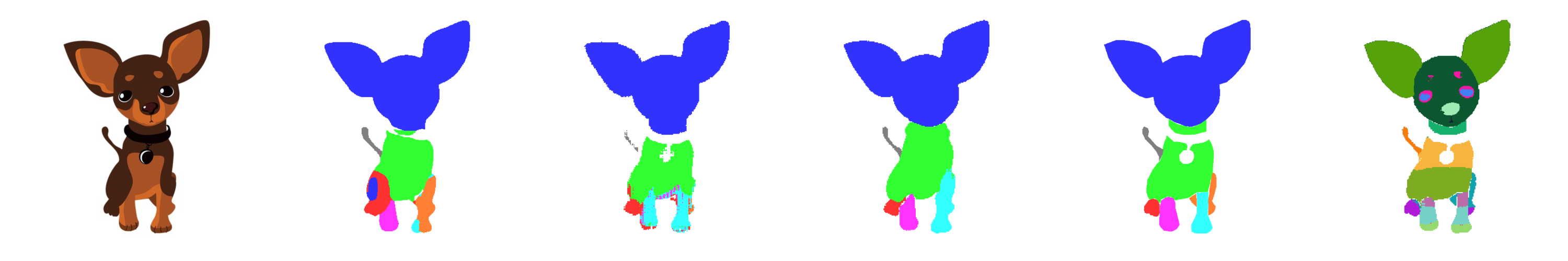}
    \includegraphics[width=\linewidth, trim={12.5cm 0 0 0},  clip]{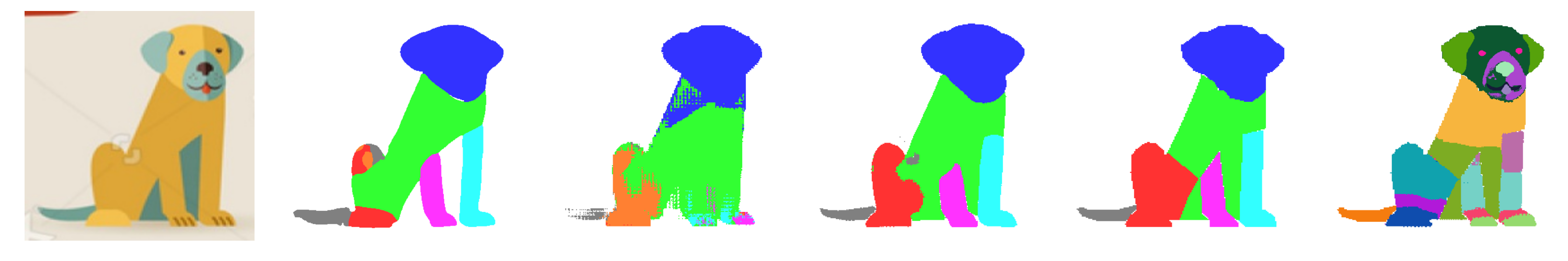}
    \caption{Cartoon dog segmentation results comparing DFPnet\cite{wan2020dense}, our model with pretrained image encoder on SA-1B(*) and on  ChildlikeSHAPES (+). Input images are omitted to avoid copyright violation. Please refer to image-ids 791, 932, 889, 807, 780, 1274 (top-to-bottom) in validation set of CartoonDogs dataset.}
    \label{fig:cartoon_dog_qual}
\end{figure}

\begin{figure}
    \centering
    \includegraphics[width=\linewidth]{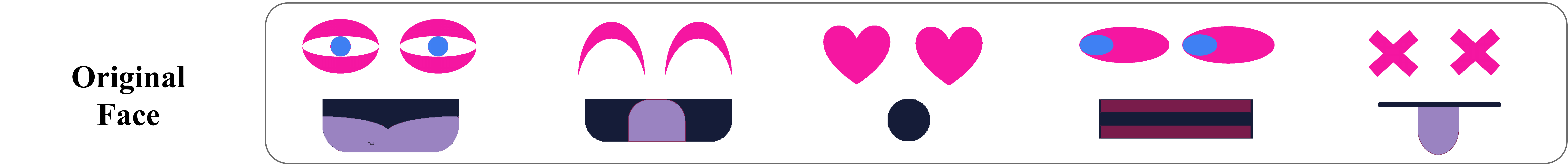}
    \includegraphics[width=\linewidth]{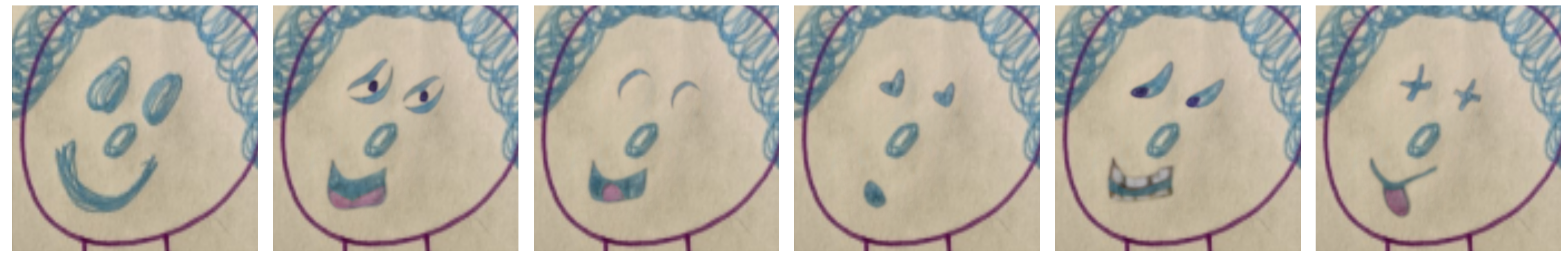}
    \includegraphics[width=\linewidth]{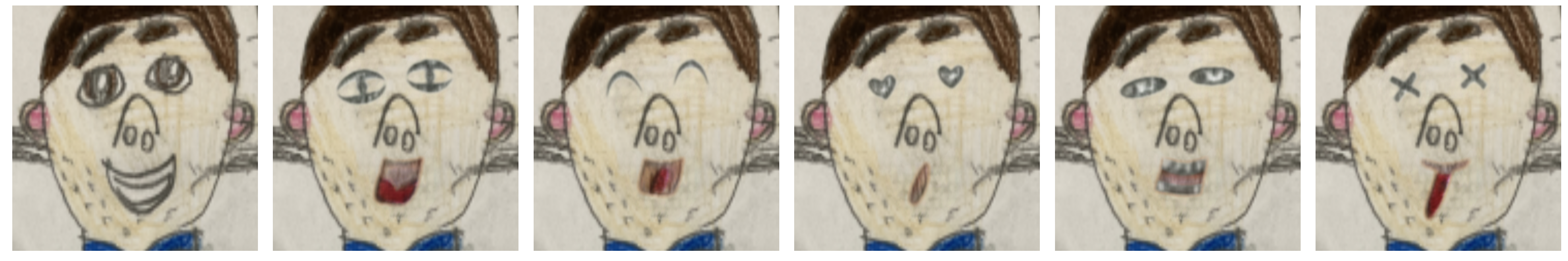}
    \includegraphics[width=\linewidth]{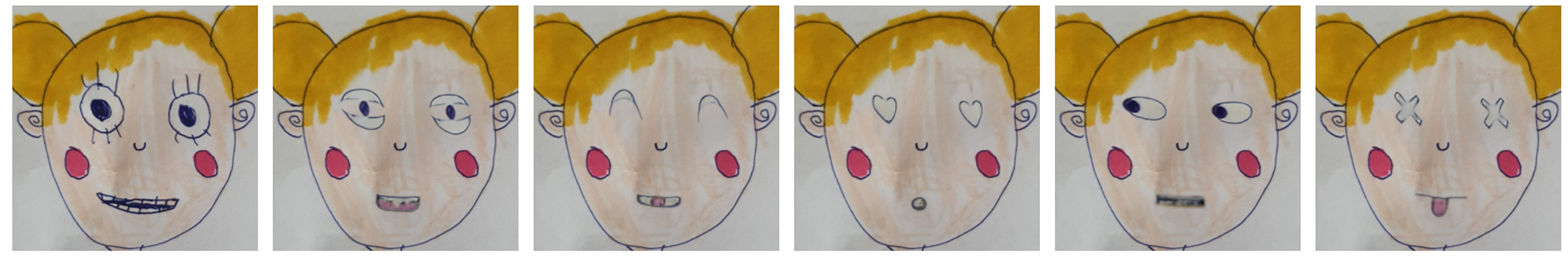}
    \includegraphics[width=\linewidth]{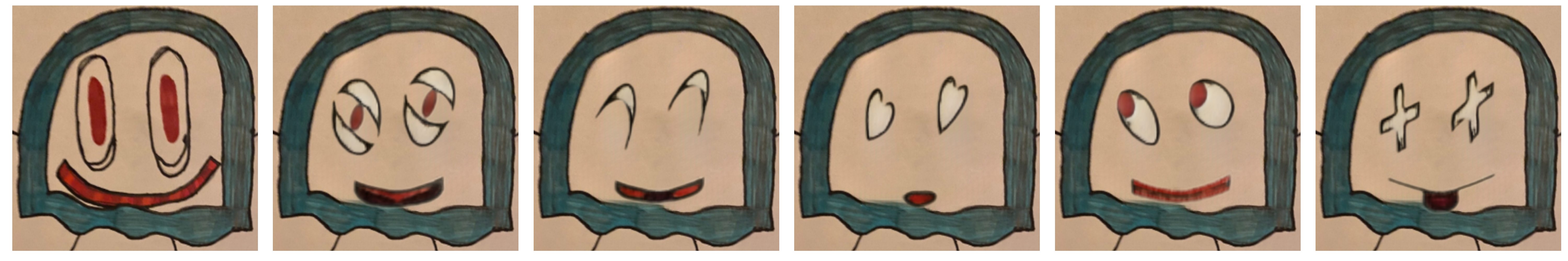}
    \includegraphics[width=\linewidth]{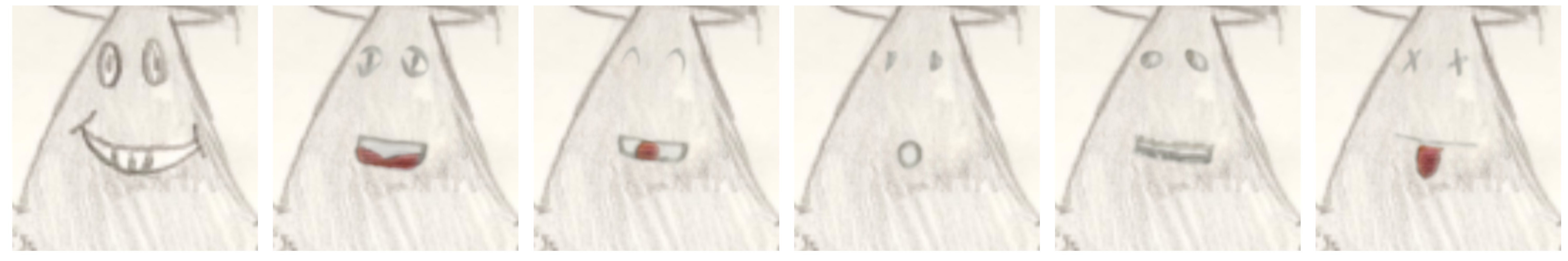}
    \includegraphics[width=\linewidth]{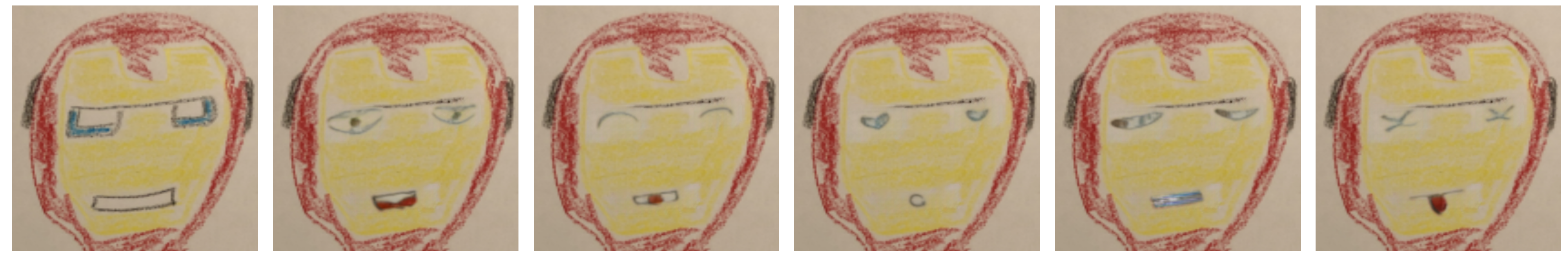}
    \caption{Results of proposed preset based facial expression generation pipeline on diverse drawing styles.}
    \label{fig:face_exp_gen_results}
\end{figure}

\begin{figure}
    \centering
    \includegraphics[width=\linewidth]{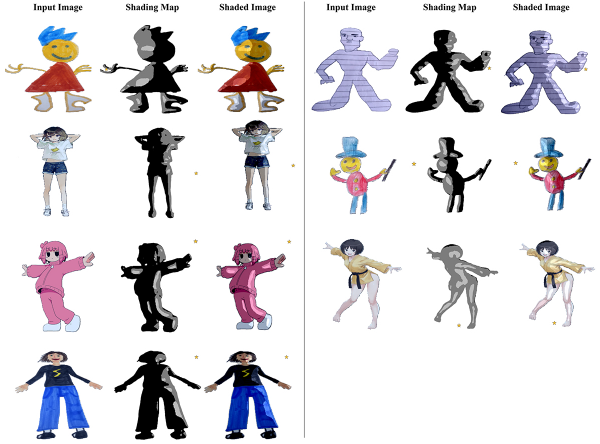}
    \caption{Automatic shading results showing input drawings, predicted shading maps, and final rendered figures. Yellow stars indicate light positions. Please refer to supplementary for video results of shading.}
    \label{fig:shading_qual}
\end{figure}


\begin{table}[h!]
    \centering
    \caption{Semantic segmentation accuracy on the ChildlikeSHAPES test set. \textit{Top:} comparison with state-of-the-art artlines \textit{Middle:} ablation studies examining the impact of encoder training and hierarchical prediction. \textit{Bottom:} our complete model with both components.}
    \label{tab:comparisons_ablations}    
    \scalebox{0.78}{
    \begin{tabular}{lccc}
        \toprule
        Method & Pixel Acc. (\%) & Mean Acc.(\%) & Mean IoU (\%)\\ \midrule
        Sapiens\cite{sapien}& 90.69 & 54.62 & 43.59 \\
        DFPnet\cite{wan2020dense}& 90.85 & 57.26 & 46.66 \\
        Mask2Former\cite{cheng2021mask2former}& 90.88 & 53.33 & 44.16 \\ \midrule
        CharSegNet w/o Encoder Training & 90.18 & 49.10 & 37.94 \\
        CharSegNet w/o Hierarchy & 91.00 & 51.41 & 40.86 \\ \midrule
        CharSegNet (Full Model) & \textbf{93.73} & \textbf{65.78} & \textbf{54.78} \\ \bottomrule
    \end{tabular}
    }

\end{table}

\begin{table}[h!]
    \centering
    \caption{Semantic segmentation accuracy on the Cartoon dog dataset. CharSegNet variants differ in initialization: directly using SAM weights (*)  versus fine-tuning on the ChildlikeSHAPES dataset. CartoonNet results are taken from the original paper.}
    \label{tab:cartoon_dog}   
    \scalebox{0.78}{    
    \begin{tabular}{lccc}
        \toprule
        Method & Pixel Acc. (\%)& Mean Acc. (\%)& Mean IoU (\%)\\ \midrule
        DFPnet\cite{wan2020dense} & 93.09 & 77.67 & 66.58 \\
        CartoonNet\cite{10.1145/3664647.3680879} & 94.64 & 84.01 & 74.05 \\
        CharSegNet* & 88.09 & 72.41 & 63.88 \\
        CharSegNet & \textbf{95.11} & \textbf{84.33} & \textbf{77.38} \\  \bottomrule
    \end{tabular}
    }

\end{table}



\section{Results}
\label{sec:results}

We present visual results demonstrating both CharSegNet's semantic segmentation capabilities and the applications they enable. We first show qualitative comparisons against baseline methods and ablations. We then show results on out-of-domain (OOD) figures and cartoon dogs. Finally, we demonstrate several applications enabled by semantic labels: novel face asset generation (Sec.~\ref{sec:facial_animation_lip_sync}), improved body animation (Sec.~\ref{sec:body_animation_improvements}), and figure shading (Sec.~\ref{sec:relighting}).

Fig.~\ref{fig:comparison_seg_all} demonstrates several advantages of CharSegNet of baselines and ablations. Importantly for downstream animation applications, our model more accurately identifies small regions like facial features (row 1). When CharSegNet's encoder is fine-tuned, it shows clear improvement upon figures with hollow body parts (rows 2, 3). CharSegNet also correctly identifies and labels overlapping body parts (row 3), a task where all baselines and ablations struggle. The benefits of the hierarchical approach are evident when compared to single-stage prediction, which can misplace body parts (e.g. feet near top in rows 1, 4) and shows greater confusion with background marks (rows 3). We provide additional results, including failure cases, in Appendix B.

Fig.~\ref{fig:comparison_ood} compares OOD results across three approaches: Sapiens fine-tuned on our dataset, an ablation of our model initialized with random weights instead of SAM's pre-trained encoder, and our full model as presented. While fine-tuning Sapiens on our dataset compromises its performance on realistic humans (rows 2-3), CharSegNet shows plausible performance across both realistic and non-realistic figures, demonstrating robust generalization to OOD samples. Sapiens frequently confuses semantic labels, such as upper leg and upper torso (rows 1 and 4), and both Sapiens and our ablated model struggle to correctly differentiate subjects from backgrounds (rows 1-3, 5). The value of SAM's pre-trained encoder weights is particularly evident in region consistency; without them, our ablated model produces checkerboard-like patterns of different labels within what should be unified regions (rows 3-5). This suggests that SAM's pre-trained weights provide a strong foundation that supports generalizing beyond childlike drawings while maintaining accurate semantic understanding. Additional examples are shown in Appendix B.

\begin{figure}
    \centering
    \includegraphics[width=0.25\linewidth]{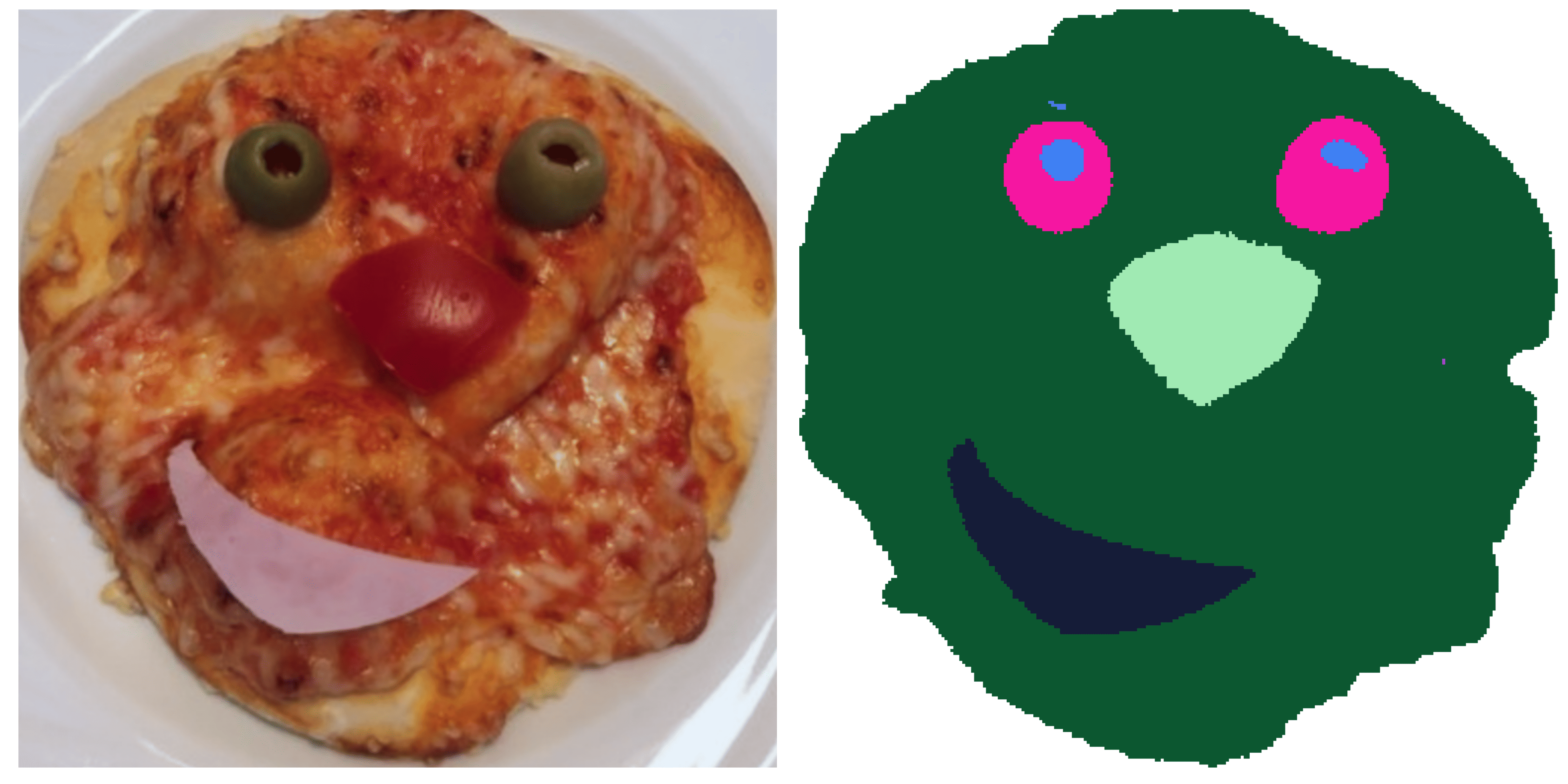}
    \includegraphics[width=0.25\linewidth]{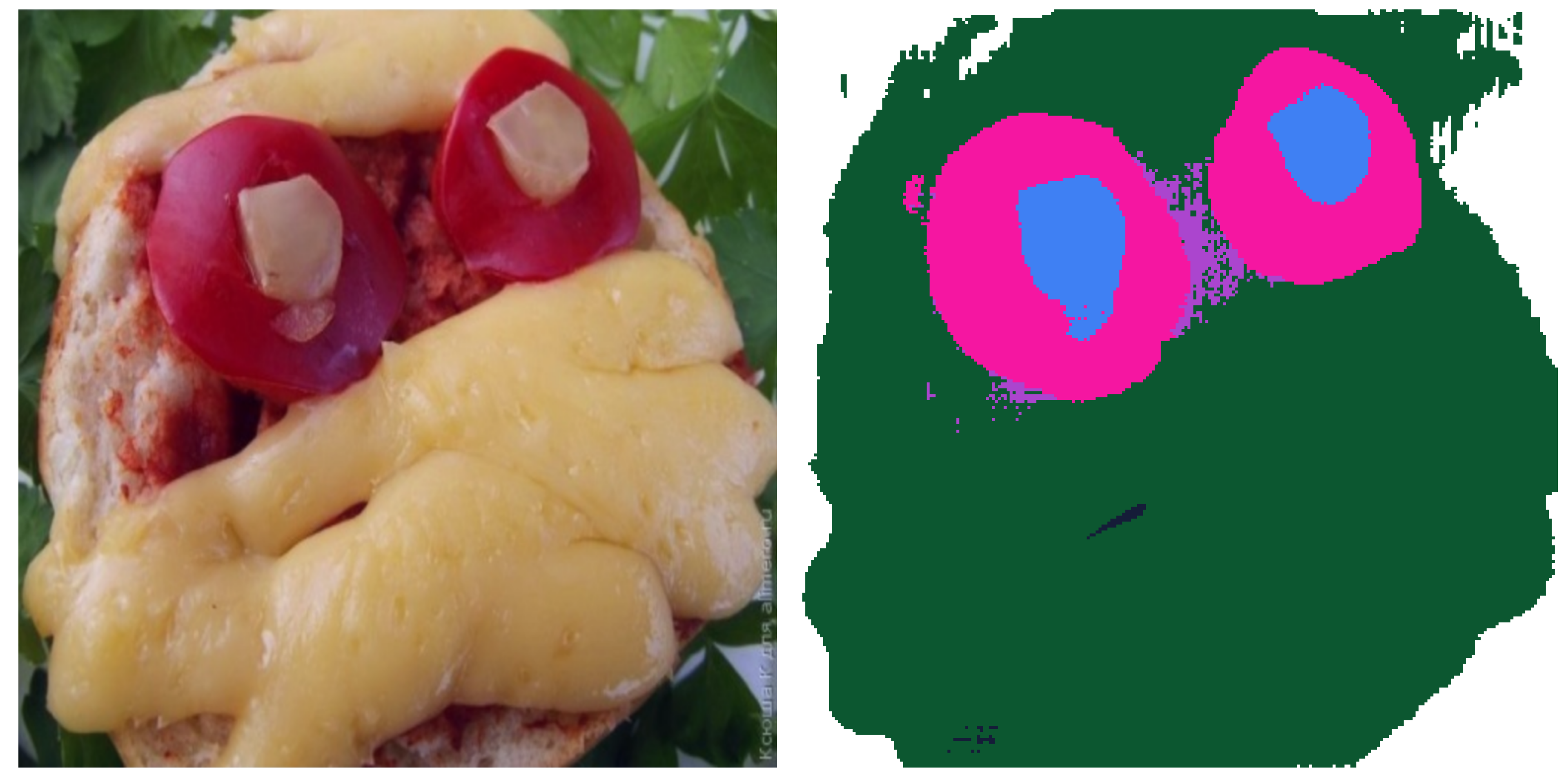}
    \includegraphics[width=0.25\linewidth]{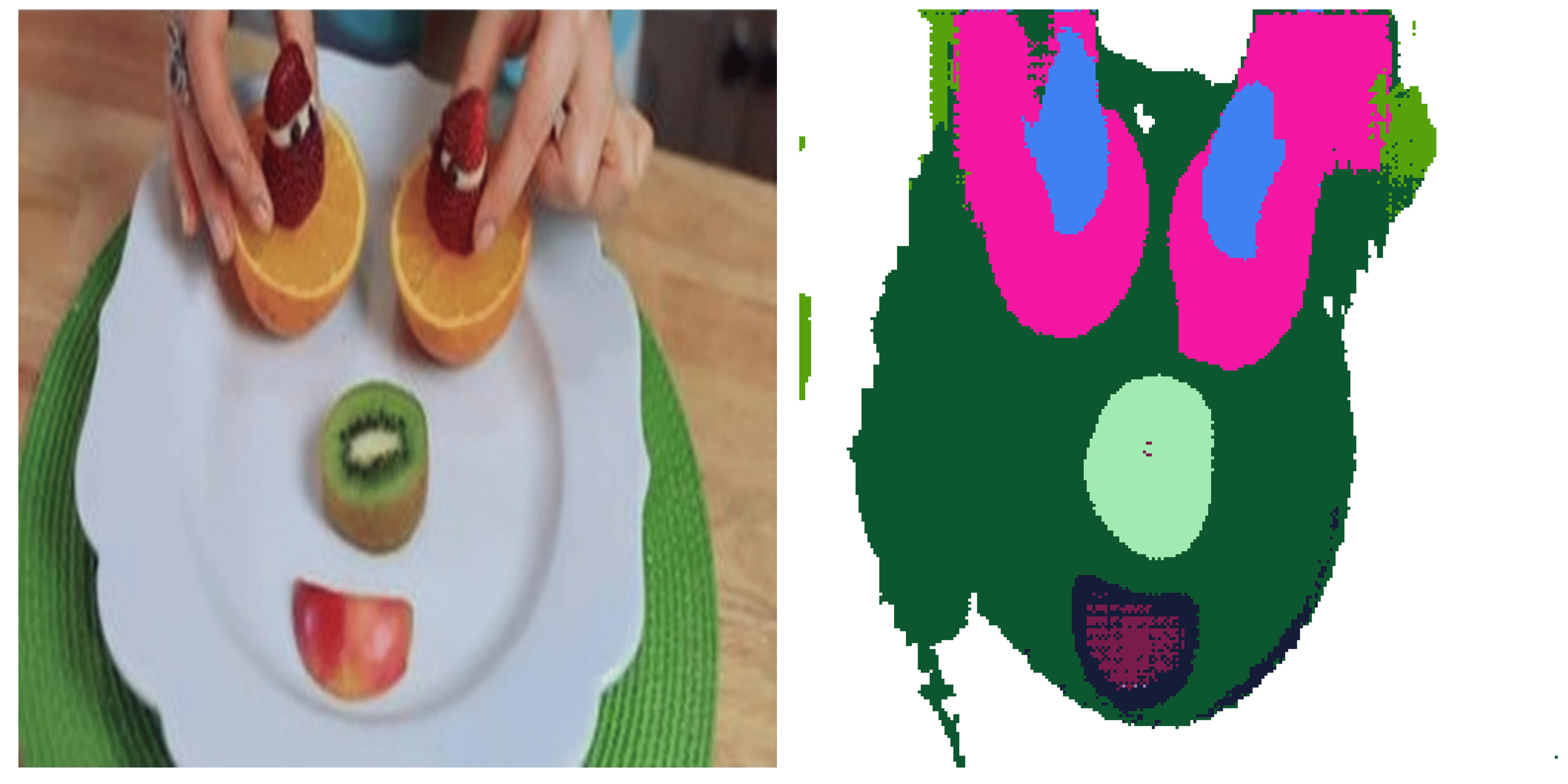}
    \includegraphics[width=0.25\linewidth]{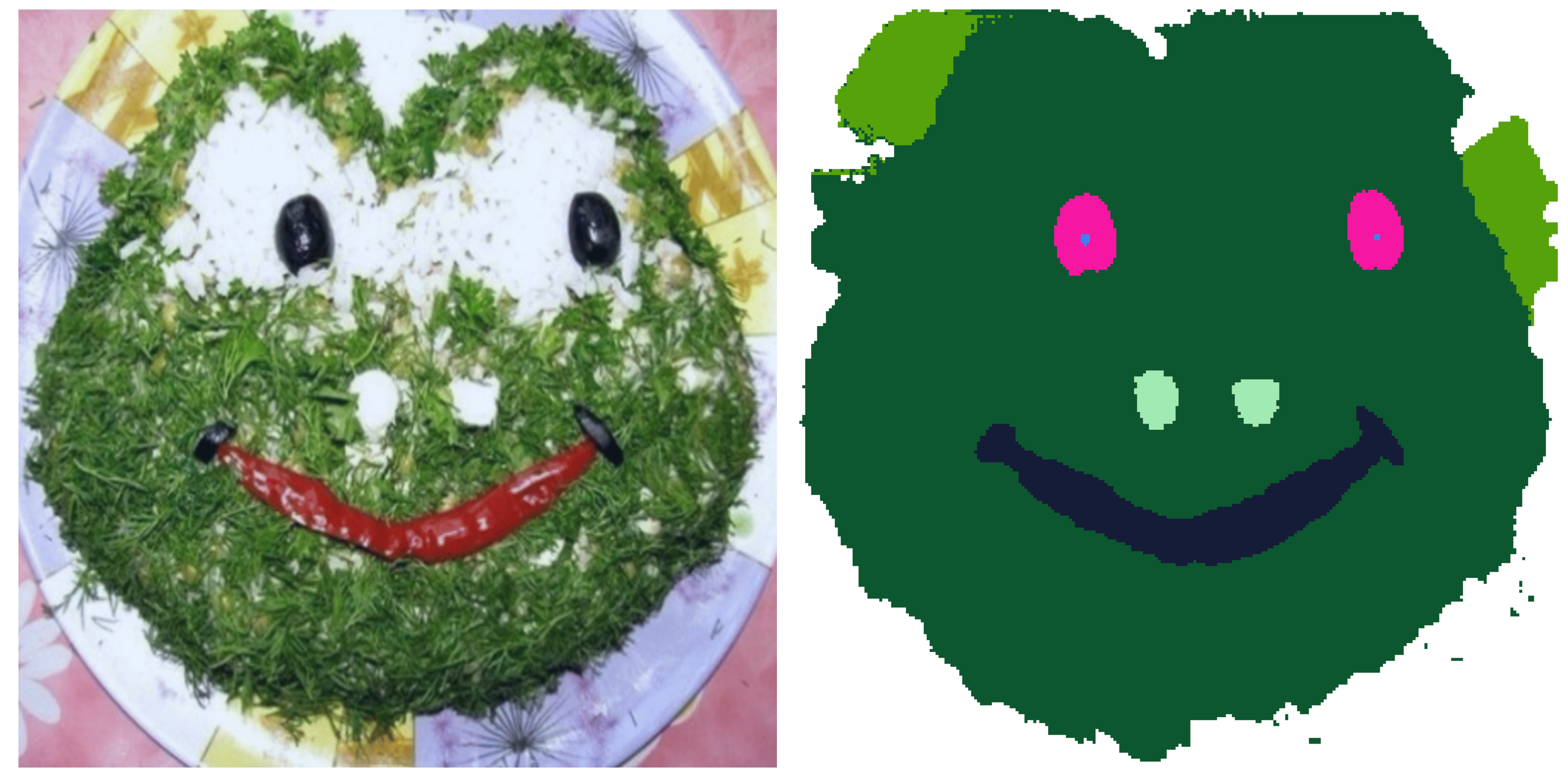}
    \includegraphics[width=0.25\linewidth]{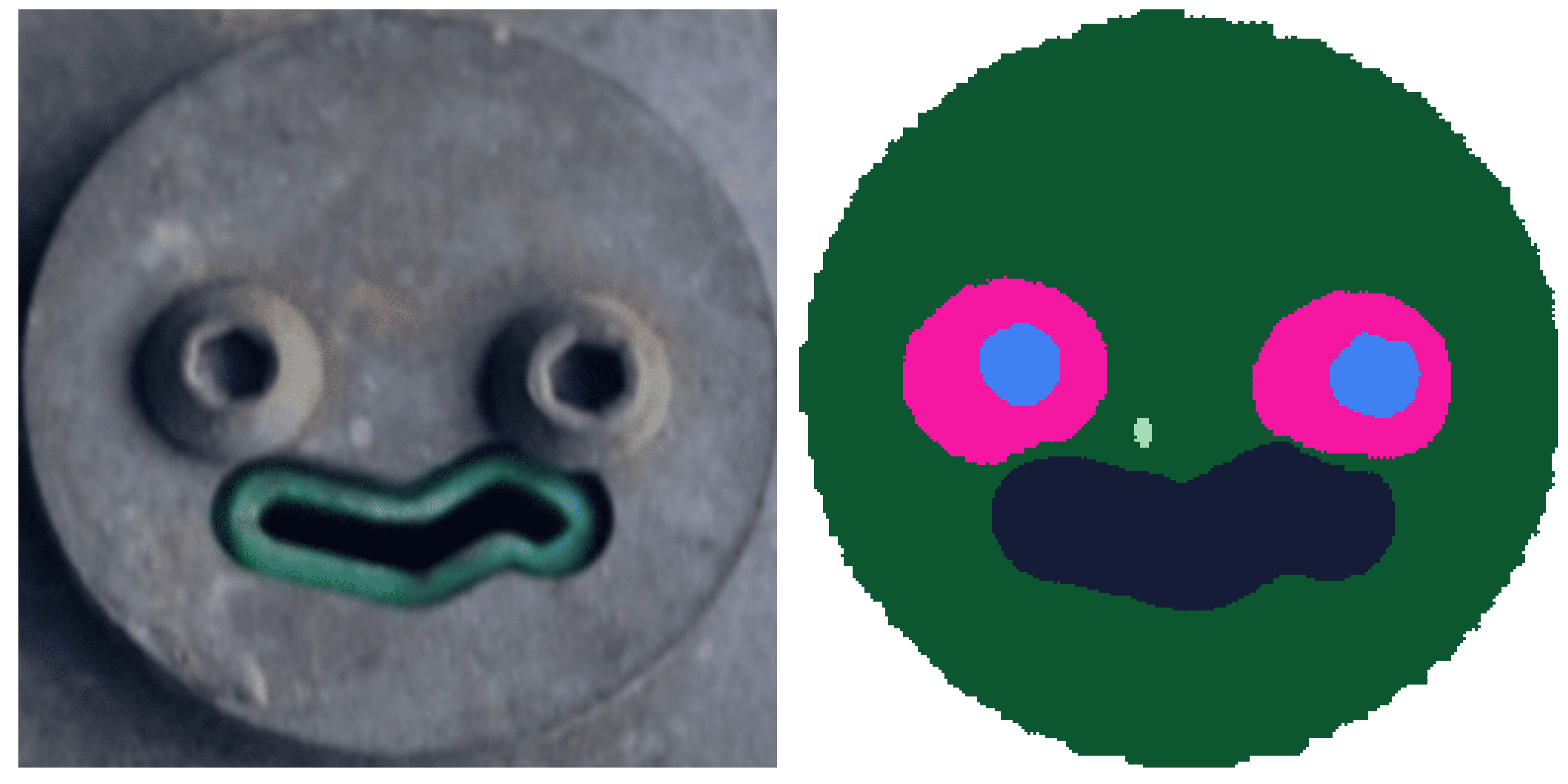}
    \includegraphics[width=0.25\linewidth]{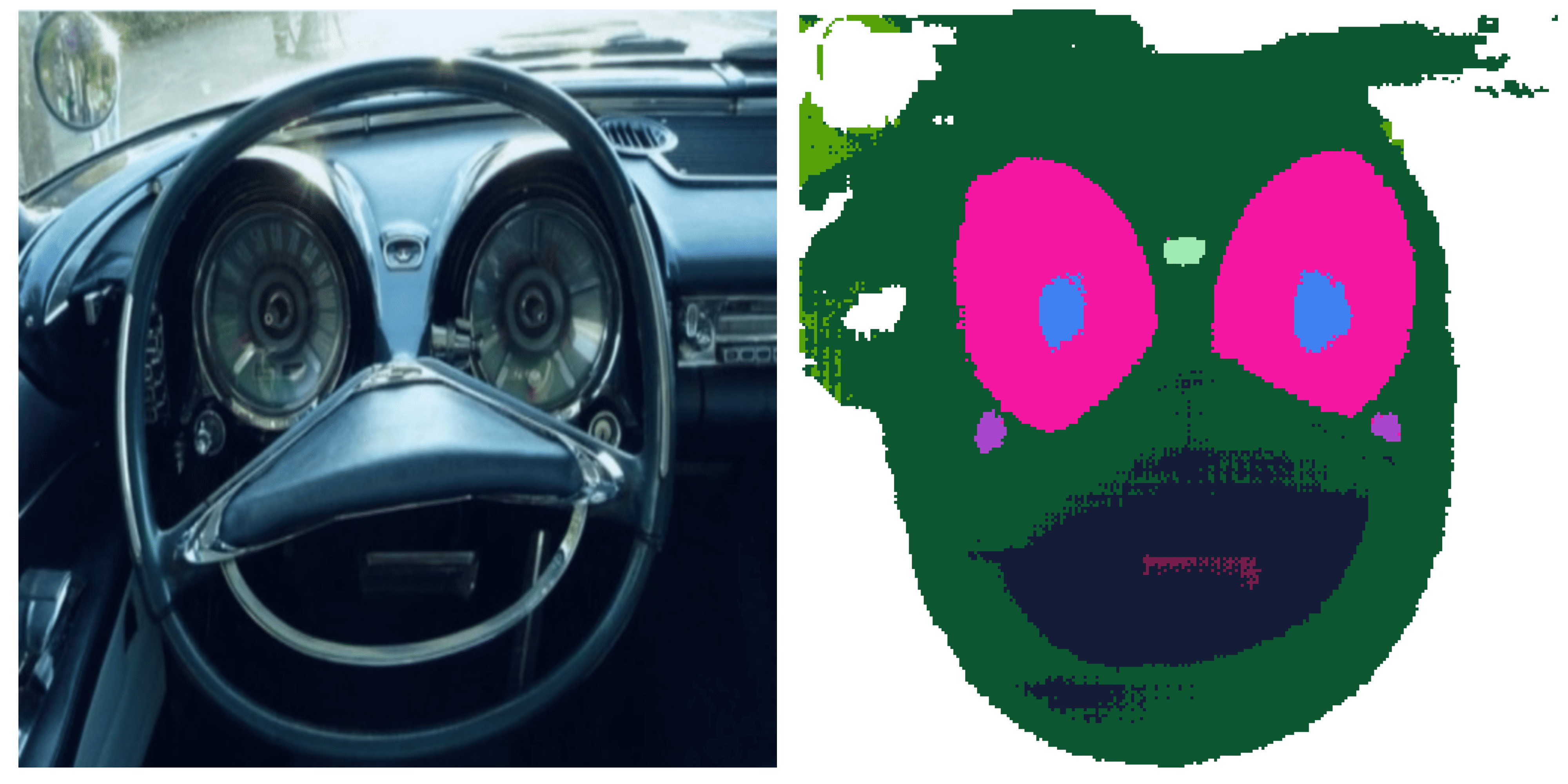}
    \includegraphics[width=0.25\linewidth]{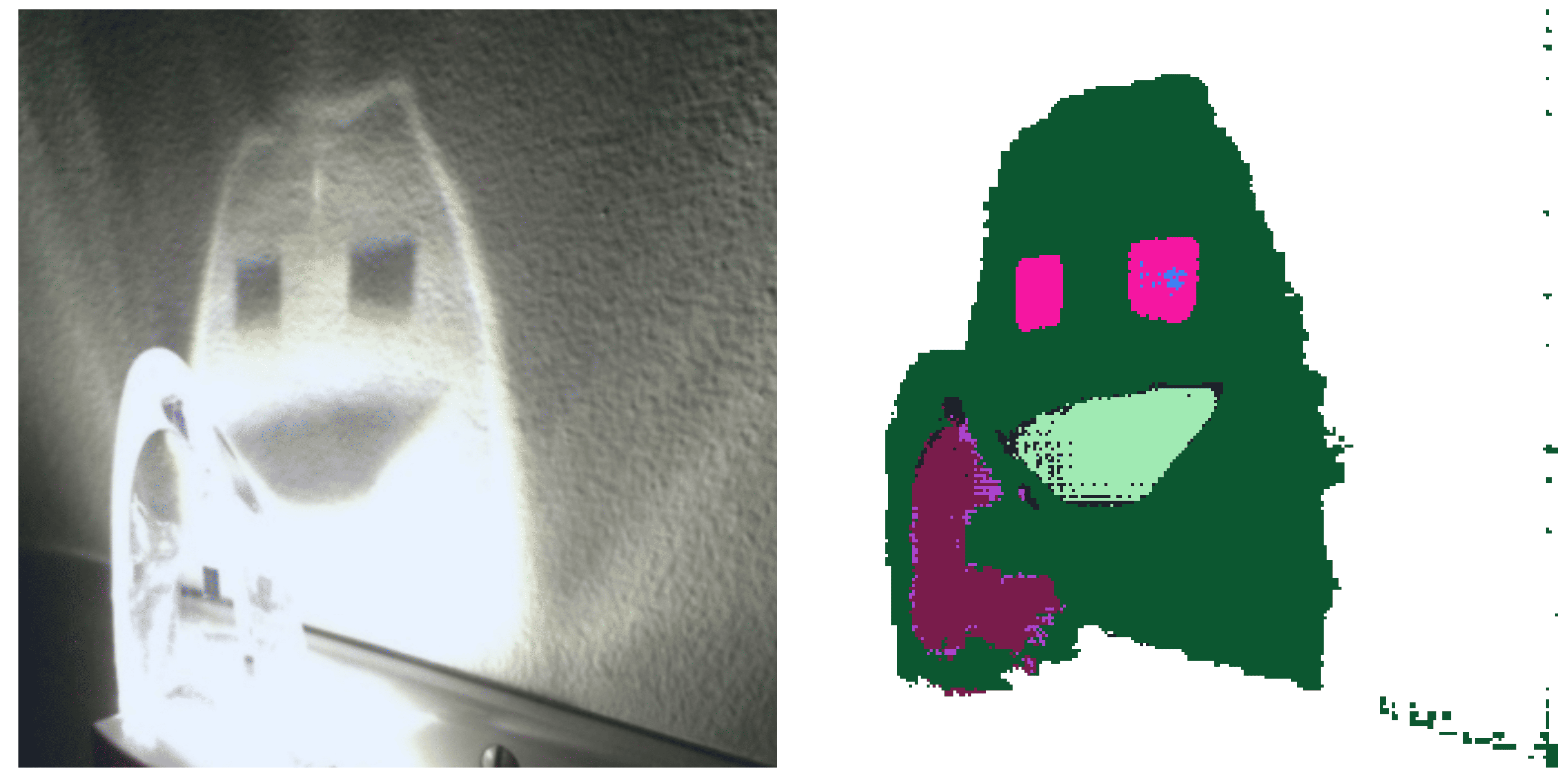}
    \includegraphics[width=0.25\linewidth]{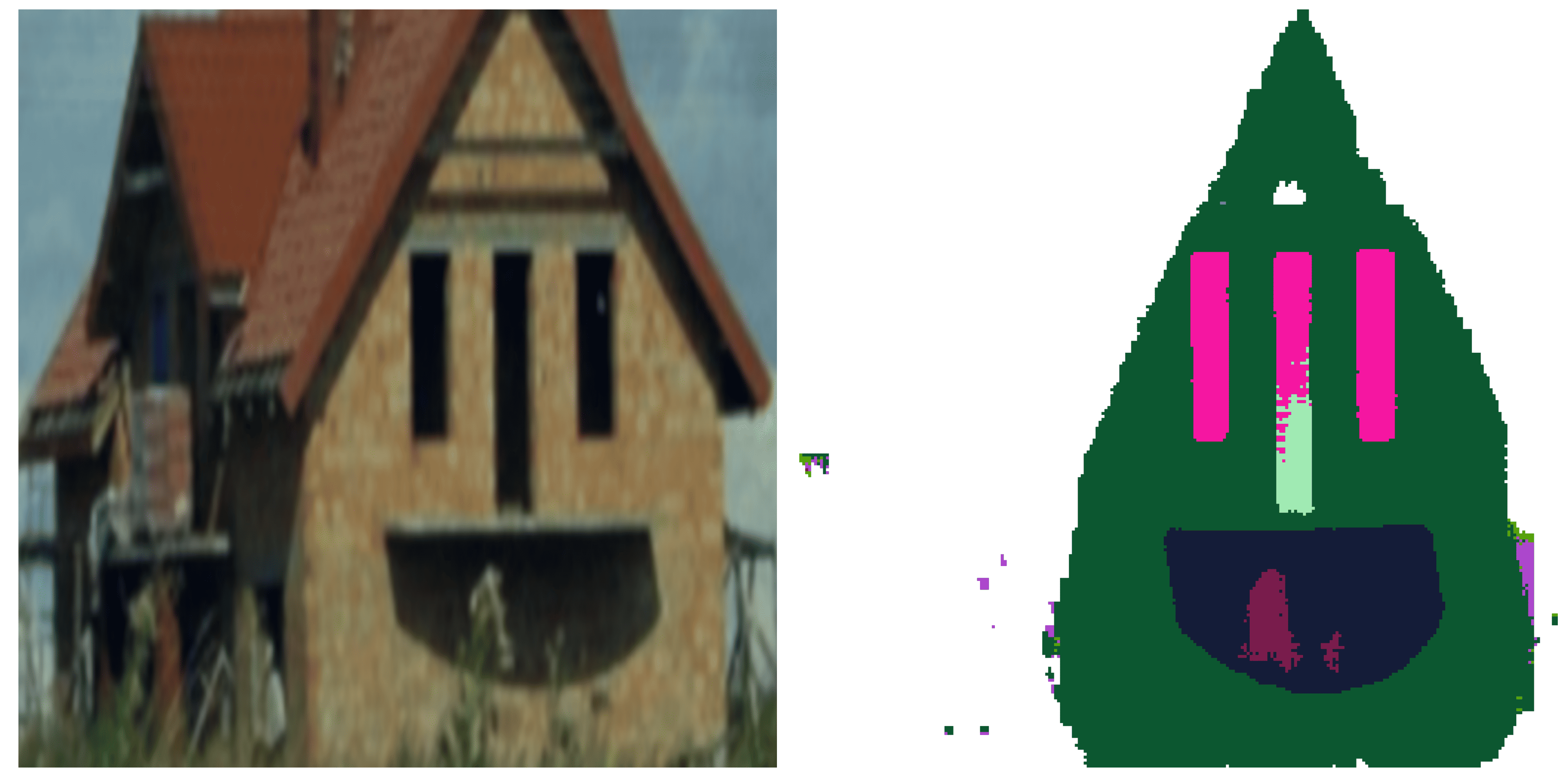}
    \includegraphics[width=0.25\linewidth]{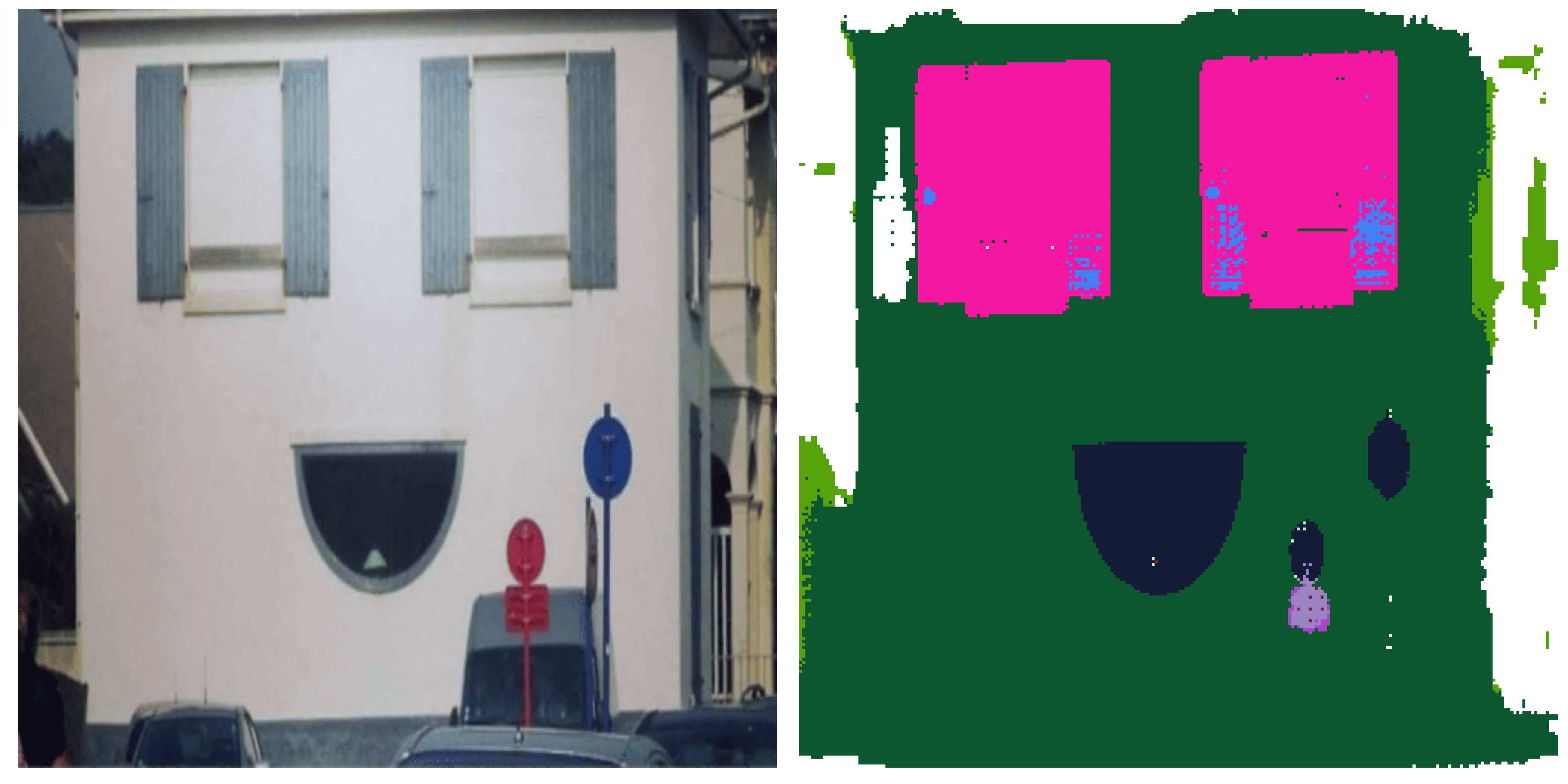}
\caption{Results of CharSegNet(Face) on images from Faces-In-Things dataset \cite{10.1007/978-3-031-73650-6_22}.}
    \label{fig:faces_in_things}
\end{figure}

Fig.~\ref{fig:out_of_domain} provides additional examples of CharSegNet's OOD performance across a diverse range of domains, including cave paintings, cubist artwork, street signs, toys and more. 
We push the limitations of our model even further by testing it up the Faces-In-Things dataset~\cite{10.1007/978-3-031-73650-6_22} (Fig.~\ref{fig:faces_in_things}).
Although not trained on these domains,
our model successfully identifies and labels semantic regions that plausibly coincide with human perception.

Fig.~\ref{fig:cartoon_dog_qual} compares segmentation performance on the Cartoon dog dataset between DFPnet and two variants of our model. Both variants use only the coarse segmentation predictor: one trained on cartoon dogs starting from SAM's pre-trained weights, and another trained on cartoon dogs after first fine-tuning SAM's weights on the ChildlikeSHAPES dataset. The variant fine-tuned with ChildlikeSHAPES demonstrates clear advantages over DFPnet, particularly in handling complex backgrounds (row 3), and obscuring objects (row 4). It demonstrates improved regional label consistency, avoiding local misclassifications within the torso and leg regions (rows 1, 5) that occur in DFPnet. 
Notably, training directly from SAM's weights without the intermediate ChildlikeSHAPES fine-tuning step significantly degrades performance, introducing a characteristic checkerboard pattern of inconsistent labels across all rows.
This further demonstrates how features learned from the ChildlikeSHAPES dataset can effectively generalize to novel artistic domains.

Fig.~\ref{fig:face_exp_gen_results} demonstrates our novel facial asset generation results (see Section ~\ref{sec:facial_animation_lip_sync}) across a variety of styles and preset templates. The mouth and eye assets generated by our CharGAN preserve the artistic style of each character while maintaining semantic correctness. CharSegNet's semantic understanding and CharGAN's style understanding work together to enable these assets to adapt appropriately to each character's unique facial structure and style. 

Fig.~\ref{fig:shading_qual} presents input images, predicted shading maps, and final shaded results achieved with our figure shading pipeline (Sec.~\ref{sec:relighting}). Despite having no explicit 3D representations of the figures, our approach produces convincing lighting effects that suggest natural volume and respond appropriately to different light positions.

\begin{figure}
    \centering
    \includegraphics[width=\linewidth]{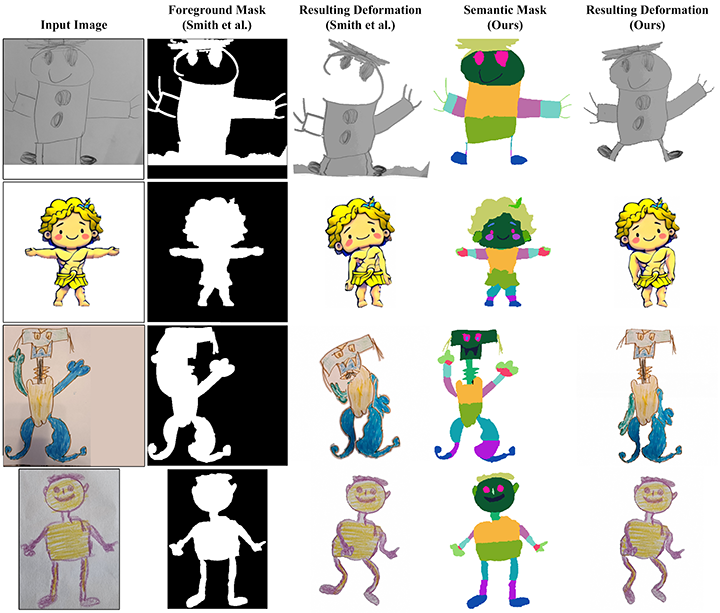}
    \caption{Visualization of how the body animation method of Smith et al. can be improved with our model's predicted semantic labels. As described in Section \ref{sec:body_animation_improvements}, we demonstrate four improvements: robust foreground segmentation (top row), part-aware deformation using modified ARAP weights (second row), separation of overlapping body parts (third row), and corrected foot orientation (last row).}
    \Description{Body Animation Results}
    \label{fig:body_animation_results}
\end{figure}

Finally, Fig.~\ref{fig:body_animation_results} demonstrates how semantic understanding can enhance existing animation techniques. By incorporating our predicted semantic labels into Smith et al.'s~\cite{10.1145/3592788} method, we improve character foreground segmentation, deformation control, part separation, and anatomical consistency.
\begin{figure}
    \centering
    \includegraphics[width=\linewidth]{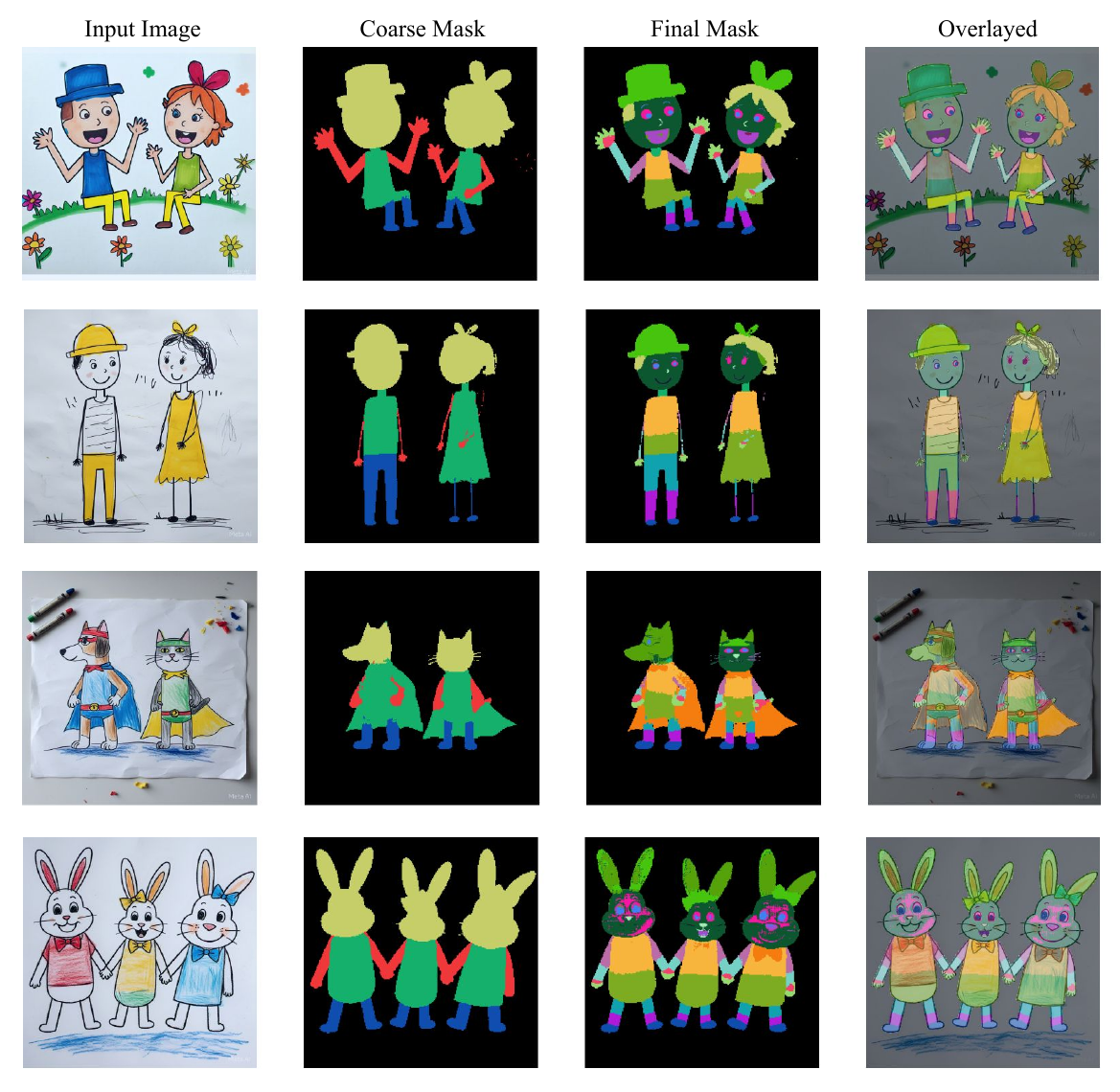}
    \caption{CharSegNet produces plausible semantic masks for multiple characters in the same image, even when not trained on such examples. This highlights the capability of our method to inherently group related regions together and employing hierarchical prediction to segment individual groups.}
    \label{fig:multichar}
\end{figure}

\begin{figure}
    \centering
    \includegraphics[width=\linewidth]{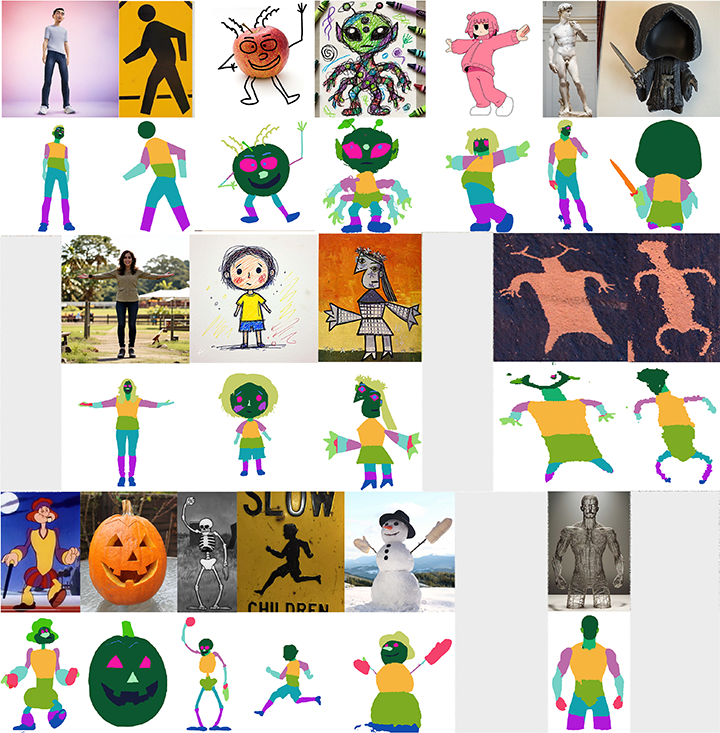}
    \caption{Semantic segmentation results on diverse out-of-domain figures, including cave paintings, cubist artwork, street signs, toys, and more.}
    \Description{Semantic segmentation example of out-of-domain human-like figures}
    \label{fig:out_of_domain}
\end{figure}

\section{Conclusion}
We have presented three main contributions: CharSegNet, a hierarchical semantic segmentation model specifically designed for childlike figure drawings; the ChildlikeSHAPES dataset, a diverse collection of 16,075 childlike drawings manually annotated with 25 classes of pixel-level semantic regions; and several novel applications demonstrating how semantic understanding can enhance character animation and create more accessible hand-drawn character animation tools. Despite only being trained on childlike drawings, our method shows remarkable generalization to out-of-domain figures, from cave paintings and snowmen to abstract art, leveraging the rich visual features of SAM's pre-trained encoder. We demonstrate this transferability further by achieving state-of-the-art results on cartoon dog segmentation, showing how features learned from childlike drawings can effectively transfer to new domains. Taken together, these results suggest that our model has learned meaningful semantic features that can benefit a range of applications beyond hand-drawn figure animation.

Our results suggest that CharSegNet has learned fundamental patterns in how humans represent figures through drawing. Unlike photographs, which inherently contain texture, color, and realistic proportions, childlike drawings capture what humans consider essential about figure representation - a form of visual abstraction that appears across cultures and throughout history. Our model appears to have internalized these representational strategies, developing a geometric understanding of how humans typically arrange and represent body parts in 2D space. To further support this conclusion, we provide a visualization of the predicted semantic heatmaps in Appendix D.

This geometric understanding strongly influences predictions outside the domain of front-facing, childlike figure drawings. When presented with quadrupeds, the model splits torso regions horizontally to maintain its learned spatial arrangement of upper and lower torso (Fig.~\ref{fig:cartoon_dog_qual}, row 1). When analyzing people in profile, it often labels ears as eyes because their rounded shapes and centered placements better align with eyes in childlike facial drawings. When presented with foreshortened limbs or partial figures, it maintains childlike drawing conventions by splitting regions at consistent relative proportions, interpreting fore-shorted arms as short arms and truncated legs as complete but diminutive legs. The model's reliance on geometric context also explains its difficulty with cropped facial images- without the broader body context, it struggles to correctly identify a disembodied face as such, instead searching within it for a full figure. For examples of all of these, see Appendix D. This behavior demonstrates how deeply CharSegNet has learned the representational principles of childlike drawings, explaining both its effective transfer to domains that share this abstract form of figure representation as well as its predictable failures when presented with subject's whose arrangements differ significantly from childlike drawing conventions.

The ChildlikeSHAPES dataset reveals several inherent ambiguities in how childlike drawings represent human figures. First, many regions permit multiple valid semantic interpretations- a triangular lower body might reasonably be labeled either as a skirt or a torso, and a shirt sleeve could be considered part of either a torso or arm, depending on the desired end effect. Because the dataset only contains one set of annotations per figure, it does not capture this range of valid interpretations. Second, overlapping regions, such as arms drawn across the torso or legs extending into the lower torso, inherently belong to multiple semantic classes but are constrained to single labels in our annotation scheme. Finally, the informal nature of childlike drawings often results in scribbled or imprecise regions where semantic boundaries become truly difficult to define. These fundamental ambiguities not only affect our evaluation metrics, but also suggest that future work should explore formulating childlike segmentation as a multi-label problem.

Though here we focus on images containing a single figure, CharSegNet is robust to images with multiple characters (see Fig.~\ref{fig:multichar}). Future work might focus on tools for automatically animating drawing scenes comprised of multiple figures, further elevating their creative potential.

These challenges invite broader questions about how to formulate and study childlike figure drawings. While multi-label approaches could address current annotation limitations, the fundamental nature of these drawings- their universality, their distillation of essential figure characteristics- suggests value far beyond computer graphics applications. The ChildlikeSHAPES dataset could provide insights for neuroscientist studying how the brain processes and recognizes abstract figures, cognitive scientists examining core principles of human figure perception, machine learning researchers exploring robust recognition and transfer learning across different representational domains, and educators developing creative tools for children. Combined with our technical approach, it opens new possibilities for accessible creative tools and creative expression. As research continues to explore how humans represent and perceive human figures, childlike drawings may prove a valuable bridge between human and machine perception.

\bibliographystyle{ACM-Reference-Format}
\bibliography{bibliography}

\end{document}